\documentclass[a4paper,11pt]{article}
\usepackage{fullpage}
\usepackage[latin1]{inputenc}
\usepackage{graphicx, graphics, epsf, epsfig}
\usepackage{indentfirst}
\usepackage[T1]{fontenc}
\usepackage{amsmath}
\usepackage{amssymb}
\usepackage{amsfonts}
\usepackage{latexsym}
\usepackage{fancybox}
\usepackage{float}

\newcommand{\barr}{\begin{array}}
\newcommand{\earr}{\end{array}}
\begin{document}
\titlepage
\title{\bf Simulation of seismic response in a city-like environment\rm}
\author{
Jean-Philippe Groby\thanks{CNRS/LMA, 31 Chemin Joseph Aiguier, 
13402 Marseille cedex 20, FRANCE, ({\tt groby@lma.cnrs-mrs.fr})}
 \and Chrysoula
Tsogka\thanks{Mathematics Department, Stanford University, USA({\tt tsogka@math.stanford.edu})}
 \and Armand
Wirgin\thanks{CNRS/LMA, 31 Chemin Joseph Aiguier, 
13402 Marseille cedex 20, FRANCE, ({\tt wirgin@lma.cnrs-mrs.fr})}
}
\date{\today}
\maketitle
\begin{abstract}
We study the seismic response of  idealized 2D cities, constituted by non equally-spaced, non equally-sized  homogenized blocks anchored in a soft layer overlying a hard half space. The blocks and soft layer are occupied by dissipative media. To simulate such response, we use an approximation of the viscoelastic modulus by a low-order rational function of frequency and incorporate this approximation into a first-order-in-time scheme. Our results display spatially-variable, strong, long-duration responses inside the blocks and on the ground, which qualitatively match the responses observed in some earthquake-prone cities of Mexico, France, the USA, etc. 
\end{abstract}
\section{Introduction}
A noticeable feature of many earthquake-prone cities such as Mexico-City \cite{baee88}, \cite{chba94}, \cite{fasu94}, Nice \cite{sedu00}, Los Angeles \cite{ol00-7}~ etc., is that they are partially or wholly built on soft soil. Seismogram records  in such cities indicate amplification of the ground motion, beating phenomena, long codas and substantial spatial variability of response.

To analyze the possible causes of these puzzling effects, we study the action of a seismic wave on a relatively-simple structural model with both geological and man-made features. Our 2D model has three components, as in \cite{Wir2003}, (from bottom to top in Fig. \ref{Conf}): a hard half space (HHS), overlain by a soft dissipative soil layer (SL), in which are partially imbedded a set of even softer dissipative blocks (SB). HHS and SL are geological features, the set of SB, which are homogenized for purpose of the analysis, is man-made and constitutes the visible component of an idealized city.

Real media disperse and attenuate waves. In \cite{Wir2003}, all the material components of the 'city' were taken to be lossless. In order to obtain a more realistic picture of wave propagation in the soft layer and in the homogenized blocks, we consider them to be composed of dissipative media. We follow the approach exposed in \cite{Gro2003}, \cite{grts04a}, \cite{grts04b} and show how it can be used to simulate SH viscoelastic wave propagation.

Even when the media are dissipative, we find that the presence of the blocks gives rise to ground motion  amplification, beating phenomena, long duration and  spatial variability of response, which are not present in the case of a flat ground (i.e., without homogenized blocks) for the same solicitation. We also find that these effects are similar, although with less amplitude, to those for non-dissipative media (as assumed in \cite{Wir2003}).

In order to understand, identify, and quantify some causes of these features, we study the influence of the distance separating the homogenized blocks. If the homogenized blocks are considered  to be  homogenized buildings, we show that decreasing the distance between buildings increases the interaction between buildings, and between buildings and the soft layer, but does not modify fundamentally the response in terms of amplification and duration. We attribute duration effects  to the excitation of Love-like modes due essentially to the presence of the blocks \cite{Wir1996}, \cite{wiko93}. 
\section{Basic Ingredients of our Model}\label{s1}
\begin{figure}[H]
\begin{center}
\includegraphics[width=12cm]{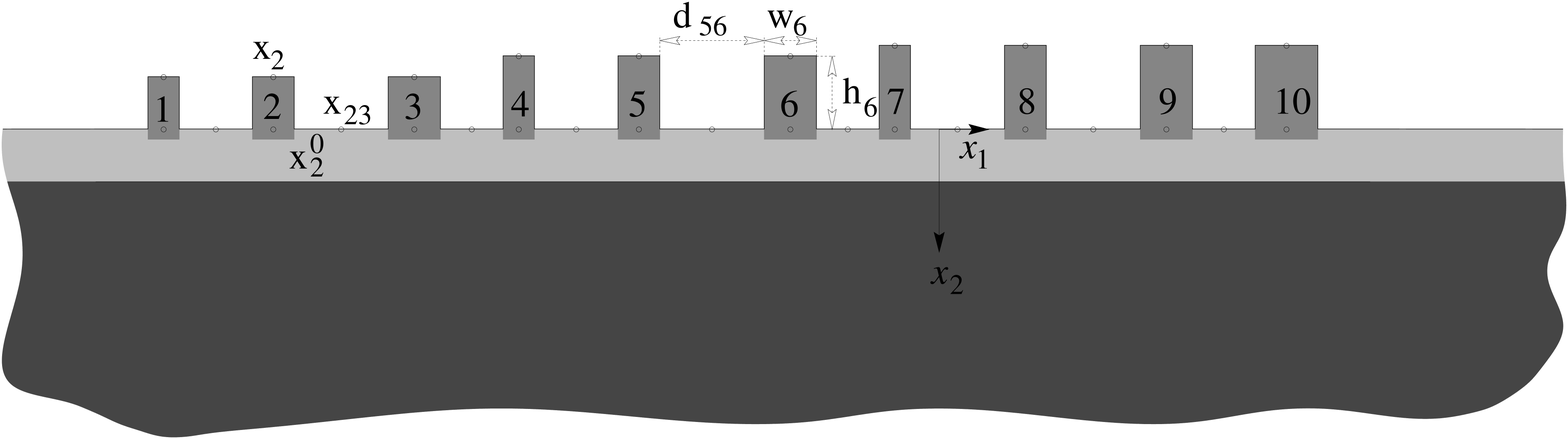}
\caption{Sagittal plane view of the city-like environment  with homogenized blocks embedded in a soft layer overlying a hard half space}
\end{center}
\label{Conf}
\end{figure}

Referring once again to Fig. \ref{Conf}, the blocks of the city are assumed to be in welded contact with the substratum by an interface on which we assume (as in \cite{Wir2003}) the continuity of displacement and normal stress. Our ``city'' is invariant in the $x_{3}$-direction with $x_{1}$, $x_{2}$, $x_{3}$ being the cartesian coordinates, and $x_{2}$ increasing with the depth (see Fig.\ref{Conf}  wherein the  sagittal plane is displayed). The support of the seismic source is a line in the $x_{3}$-direction, and is located deep in the HHS, radiating a Ricker pulse cylindrical shear-horizontal (SH) displacement field. Only the $x_{3}$-component of this field is non-vanishing and invariant with respect to $x_{3}$, so that the total field underneath and on the free surface is also SH-polarized and invariant with respect to $x_{3}$. The resulting problem is 2D with the displacement field depending only on $x_{1}$, $x_{2}$ and on time $t$.

We denote by $h_{i}$, $w_{i}$ and $d_{i,i+1}^{j}$, the height, width of the block $B_{i}$ and space interval between the blocks $B_{i}$ and $B_{i+1}$ for the three configurations, $C^{j}~;~j=1,2,3$, we studied.  The substratum $\Omega_{0}$ is occupied by a linear, isotropic, medium $M_{0}$, characterized by mass density $\rho(\mathbf{x})=\rho_{0}$, shear modulus $\mu(\mathbf{x})=\mu_{0}$, and quality factor $Q(\mathbf{x})=Q_{0}=+\infty$, with $\mathbf{x}=(x_{1},x_{2})$. The soft layer $\Omega_{1}$ is occupied by a linear, isotropic, dissipative medium $M_{1}$ characterized by mass density $\rho(\mathbf{x})=\rho_{1}$, relaxed shear modulus $\mu_{R}(\mathbf{x})=\mu_{R1}$, and quality factor $Q(\mathbf{x})=Q_{1}$. The blocks $\Omega_{2}$ are occupied by the linear, isotropic, dissipative medium $M_{2}$ characterized by mass density $\rho(\mathbf{x})=\rho_{2}$, relaxed  shear modulus $\mu_{R}(\mathbf{x})=\mu_{R2}$, and quality factor $Q(\mathbf{x})=Q_{2}$.
\section{Methods}\label{s2}
Since problems in seismology are essentially concerned with transient 
phenomena it is natural to treat them 
in the  space-time rather 
than the space-frequency framework. This also presents certain advantages from 
the numerical point of view. The obvious
approach is then to solve the second-order-in-time wave equation
(Navier equation when the medium is isotropic)  
for the displacement vector. A less-obvious, but
numerically-advantageous method \cite{Bec2001}, \cite{Bec2000}, \cite{Tso1999}, 
is to cast the wave equation into
the form of two first-order-in-time partial differential
equations, for the two unknowns constituted by the displacement
and  pseudo-velocity vectors.

In the following, we show briefly (more details are given in \cite{Gro2003}, \cite{grts04a}, \cite{grts04b})  how to do this
for motion in {\it viscoelastic}, isotropic media. Since our problem relates to 
2D, SH motion, the unknowns reduce to the single non-zero component of the
displacement vector and the two components of a pseudo-velocity
vector. We show that these unknowns satisfy two coupled first-order-in-time partial 
differential  equations, in which appear a set of unknowns (usually called memory variables) whose existence is due to 
the viscoelastic nature of the media. Each of these coefficients is shown to satisfy  a first-order-in-time partial 
differential equation whose driving term is related to the pseudo-velocity vector.

In an isotropic, non pre-stressed, elastic ( viscoelastic) medium $M$, the stress tensor $s_{kl}$ 
 is related to 
the strain tensor $e_{kl}$ by the  Hooke-Cauchy constitutive
relation 
\begin{equation}\label{2.1}
s_{kl}=\lambda\delta_{kl}e_{mm}+2\mu e_{kl}~~;~~k,l=1,2,3
 ~.
\end{equation}
wherein the Einstein summation convention is implicit,
$\delta_{kl}$ is the Kronecker delta and $e_{kl}$ is 
related to the particle displacement vector (whose elements are $u_{k}$) by
\begin{equation}\label{2.2}
e_{kl}=\frac{1}{2}\left ( u_{k,l}+u_{l,k}\right)
 ~,
\end{equation}
with $ u_{k,l}$ signifying the partial derivative of $u_{k}$ with
respect to the $l$-th cartesian coordinate $x_{l}$. Furthermore,
 $\lambda$ is the
bulk modulus and $\mu$ the rigidity. In elastic media, $\lambda$
and $\mu$ are, at most, functions of position, whereas in
viscoelastic media (i.e., with memory), ~$\mu$ depends also on
time.

The tensorial wave equation in elastic and viscoelastic media, in the context of 
linear (visco) elasticity, is 
\begin{equation}\label{2.8}
 s_{kl,k}-\rho \partial_{t}^{2} u_{l}=-\rho f_{l}~~;~~k,l=1,2,3 
  ~,
\end{equation}
wherein $\rho$ is the mass density of the medium, $f_{l}$ the $l$-th component 
of the applied force density vector, $t$  the time variable and $\partial_{t}^{2}:=
\frac{\partial^{2}}{\partial t^{2}}$. This set of partial differential equations can be
recognized to be {\it linear in terms of the displacement} $u_{l}$
since $s_{kl}$ is a linear function of $u_{l}$.

The 2D SH aspect of the problem is encompassed in the assumptions: 
$f_{1}=f_{2}=0$, $u_{1}=u_{2}=0$ and $u_{3,3}=0$.
so that the tensorial wave equation becomes 
\begin{equation}\label{2.11}
 (s_{13,1}+s_{23,2})-\rho\partial_{t}^{2}u_{3}=-\rho f_{3} ~.
\end{equation}
Let
\begin{equation}\label{2.12}
 \boldsymbol{\sigma}:=(s_{13},s_{23})~~,~~u:=u_{3}~~,~~f:=f_{3}~.
\end{equation}
Then
\begin{equation}\label{2.13}
 \boldsymbol{\sigma}=(\mu u_{3,1},\mu u_{3,2})=\mu\nabla u~,
\end{equation}
 which is the expression of the
constitutive relation for the situation corresponding to that of
2D SH motion in (visco)elastic media.

By the same token, the 
wave equation for 2D SH motion in isotropic, non pre-stressed,
(visco)elastic media becomes
\begin{equation}\label{2.16}
 \nabla\cdot(\mu\nabla u)-\rho\partial_{t}^{2}u=-\rho f ~,
\end{equation}
which is of the general form
\begin{equation}\label{2.17}
 \nabla\cdot(a\nabla U)-b\partial_{t}^{2}U+cF=0 ~.
\end{equation}

We are able to show that this equation can be obtained from the two
first-order-in-time equations 
\begin{equation}\label{2.18}
 -\partial_{t}\mathbf{V}+a\nabla{U}=0~,
\end{equation}
\begin{equation}\label{2.19}
 b\partial_{t}U-\nabla\cdot \mathbf{V}=cG~.
\end{equation}
by taking the divergence of (\ref{2.18}), the partial time derivative
of (\ref{2.19}), and subtracting the equations so obtained. This
leads to (\ref{2.17}) provided $G$ is chosen so that
\begin{equation}\label{w1.5.85u}
 F=\partial_{t}G
 ~.
\end{equation}
Thus, instead of being confronted with solving a single,
second-order-in-time partial differential equation for $U$, we must  solve 
two first-order-in-time partial differential
equations for the displacement $U$ and the pseudo-velocity
$\mathbf{V}$.

By  a viscoelastic solid  we mean a material whose rigidity at an instant $t$
depends on the whole strain history of the solid, so that 
the Hooke-Cauchy relation takes the more general (convolution)
form (for 2D, SH motion)
\begin{equation}\label{3.1}
 \boldsymbol{\sigma}(\mathbf{x},t)=\int_{-\infty}^{t}\mu(\mathbf{x},t-\tau)
 \nabla u(\mathbf{x},\tau)d\tau~,
\end{equation}
wherein $\mathbf{x}:=(x_{1},x_{2})$. When the
constitutive relation (\ref{3.1}) is inserted into the wave
equation, the latter becomes an integro-differential equation
which is very cumbersome to solve numerically because it requires
saving in memory the whole history of the solution at all points
of the compuational domain. To overcome this inconvenience,
Emmerich and Korn \cite{Emm1987} propose a (essentially Maxwell
body) method whereby they approximate the Fourier spectrum
$\mu(\mathbf{x},\omega)$ (with $\omega$ the angular frequency) of
the viscoelastic rigidity modulus $\mu(\mathbf{x},t)$ by a
rational function of frequency.

Consider  $\mathcal{F}(\mathbf{x},t)$  to be a function of $\mathbf{x}$
and $t$;  then $\mathcal{F}(\mathbf{x},\omega)$ is its Fourier spectrum (with $\omega$ the angular frequency)
such that
\begin{equation}\label{3.2}
 \mathcal{F}(\mathbf{x},t)=\int_{-\infty}^{\infty}
 \mathcal{F}(\mathbf{x},\omega)\exp(-i\omega t)d\omega~.
\end{equation}
Consequently, the frequency domain constitutive relation corresponding to
(\ref{3.1}) is
\begin{equation}\label{3.3}
 \boldsymbol{\sigma}(\mathbf{x},\omega)=\mu(\mathbf{x},\omega)\nabla u(\mathbf{x},\omega)~.
\end{equation}
We suppose that the spectrum function $\mu(\mathbf{x},\omega)$ is
known (from experimental data) and wish to approximate it in the rational function manner.
To do this, we introduce the relaxation function $R(\mathbf{x},t)$
defined by
\begin{equation}\label{3.4}
 \mu(\mathbf{x},t)=\frac{dR(\mathbf{x},t)}{dt}~.
\end{equation}
This relaxation function is assumed in \cite{Emm1987} to take the
form of a discrete sum of sinusoidal functions
\begin{equation}\label{3.5}
 R(\mathbf{x},t)\approx\mu_{R}(\mathbf{x})\left[ 1+\frac{\delta\mu(\mathbf{x})}
 {\mu_{R}(\mathbf{x})}\sum_{j=1}^{J}a_{j}(\mathbf{x})\exp(-\omega_{j}t)\right]
 H(t)~,
\end{equation}
wherein $J$ is some integer, preferably not too large to reduce
the computational effort, $H(t)$ is the Heaviside function,
$\mu_{R}(\mathbf{x})$ the relaxed rigidity modulus (equal to the
time-invariant rigidity in the elastic case),
$\delta\mu(\mathbf{x})$ a differential rigidity which is a measure
of the departure from elasticity, $\omega_{j}$ the relaxation
frequencies, it being understood that the coefficients $a_{j}$
obey the (normalization) relation
\begin{equation}\label{3.6}
\sum_{j=1}^{J}a_{j}(\mathbf{x})=1~.
\end{equation}
From (\ref{3.4})-(\ref{3.6})we find
\begin{equation}\label{3.10}
 \mu(\mathbf{x},\omega)\approx\mu_{R}(\mathbf{x})\left[ 1+
 \sum_{j=1}^{J}y_{j}(\mathbf{x})\frac{i\omega}{i\omega-\omega_{j}}\right]
~,
\end{equation}
wherein
\begin{equation}\label{3.9}
 y_{j}(\mathbf{x}):=\frac{\delta\mu(\mathbf{x})}
 {\mu_{R}(\mathbf{x})}a_{j}(\mathbf{x})
~.
\end{equation}
In the works of Emmerich and Korn \cite{Emm1987} and Groby and
Tsogka \cite{Gro2003}, \cite{grts04a}, \cite{grts04b} the weight functions
$\{y_{j}(\mathbf{x}~;~j=1,2,..,J)\}$ are determined in unique
manner from a given $\mu(\mathbf{x},\omega)$ by solving an
overdetermined linear system arising from sampling $(\ref{3.10})$
at a set of relaxation frequencies $\omega_{l}$ which are chosen
equidistant on a logarithmic scale in the interval $\left[
\omega_{max}/100, \omega_{max}\right]$, with $\omega_{max}$ the
maximal frequency of the solicitation spectrum.

We introduce the new variables
$\boldsymbol{\zeta}_{j}(\mathbf{x},\omega)$:
\begin{equation}\label{3.11}
\boldsymbol{\zeta}_{j}(\mathbf{x},\omega):=\frac{i\omega}{i\omega-
\omega_{j}}y_{j}(\mathbf{x})\mu_{R}(\mathbf{x})\nabla
u(\mathbf{x},\omega) ~,
\end{equation}
so that
\begin{equation}\label{3.20}
\boldsymbol{\sigma}(\mathbf{x},t)=\mu_{R}\nabla u(\mathbf{x},t)+\sum_{j=1}^{J} \boldsymbol{\zeta}_{j}(\mathbf{x},t)=\boldsymbol{\sigma}_{R}(\mathbf{x},t)+
\sum_{j=1}^{J} \boldsymbol{\zeta}_{j}(\mathbf{x},t)~.
\end{equation}
It is then straightforward to show that $\boldsymbol{\zeta}_{j}$ satisfies the differential equation:
\begin{equation}\label{3.18}
\partial_{t}\boldsymbol{\zeta}_{j}(\mathbf{x},t)+
\omega_{j}\boldsymbol{\zeta}_{j}(\mathbf{x},t)=
y_{j}(\mathbf{x})\partial_{t}\boldsymbol{\sigma}_{R}(\mathbf{x},t)~.
\end{equation}

We now introduce a new function
$\eta_{j}(\mathbf{x},t)$ such that
\begin{equation}\label{3.22}
 \rho(\mathbf{x})\partial_{t}^{2}\eta_{j}(\mathbf{x},t):=
 \nabla\cdot\boldsymbol{\zeta}_{j}(\mathbf{x},t)~.
\end{equation}
Then, it can be shown that:
\begin{equation}\label{3.26}
\partial_{t}^{2}\eta_{j}(\mathbf{x},t)+ \omega_{j}\partial_{t}
\eta_{j}(\mathbf{x},t)=\frac{1}{\rho(\mathbf{x})}
 \nabla\cdot\left ( y_{j}(\mathbf{x})
 \mu_{R}(\mathbf{x})\nabla u(\mathbf{x},t)\right) ~,
\end{equation}
\begin{equation}\label{3.28}
 \nabla\cdot\left( \mu_{R}(\mathbf{x})\nabla u(\mathbf{x},t)\right) +
 \rho(\mathbf{x})\sum_{j=1}^{J}\partial_{t}^{2}\eta_{j}(\mathbf{x},t)-
 \rho(\mathbf{x})
 \partial_{t}^{2}u(\mathbf{x},t)+\rho(\mathbf{x})
 f(\mathbf{x},t)=0~.
\end{equation}
The latter equation is of the form (\ref{2.17}) provided we make the
associations: $U=u$, $a=\mu_{R}$, $b=\rho$,
$F=f+\sum_{j=1}\partial_{t}^{2}\eta_{j}$, and $c=\rho$, so that
employing the previously-demonstrated equivalence between
(\ref{2.17}) and a system of two first-order-in-time partial
differential equations, we find, at present, that (\ref{3.28}) is
equivalent to the couple of equations:
\begin{equation}\label{3.29}
 \mu_{R}(\mathbf{x})\nabla u(\mathbf{x},t)-
 \partial_{t}\mathbf{V}(\mathbf{x},t)=0~,
\end{equation}
\begin{equation}\label{3.30}
 \rho\partial_{t}u(\mathbf{x},t)-\nabla\cdot\mathbf{V}(\mathbf{x},t)=
 \rho g(\mathbf{x},t)+\sum_{j=1}^{J}\partial_{t}\eta_{j}(\mathbf{x},t)~,
\end{equation}
wherein $g$ is such that
\begin{equation}\label{3.31}
 \partial_{t}g(\mathbf{x},t)=f(\mathbf{x},t)~.
\end{equation}
The last step is to introduce (\ref{3.29}) into (\ref{3.26})
followed by integration over $t$ and neglect of the integration
constant:
\begin{equation}\label{3.33}
\partial_{t}\eta_{j}(\mathbf{x},t)+ \omega_{j}\eta_{j}(\mathbf{x},t)=
\frac{1}{\rho(\mathbf{x})}
 \nabla\cdot\left ( y_{j}(\mathbf{x})\mathbf{V}(\mathbf{x},t)\right)
 ~.
\end{equation}

Our procedure for solving a
problem of 2D SH wave motion, in response to the solicitation
$f(\mathbf{x},t)$, in the temporal interval $[0,T]$ and spatial
domain $\Omega$, occupied by a isotropic, viscoelastic medium $M$
characterized by the density $\rho(\mathbf{x})$ and rigidity
spectrum function $\mu(\mathbf{x},\omega)$, thus boils down to:
\begin{itemize}
\item obtain $g(\mathbf{x},t)$ from $f(\mathbf{x},t)$ via
\begin{equation}\label{3.34}
 \partial_{t}g(\mathbf{x},t)=f(\mathbf{x},t)~~;~~
 \mathbf{x}\in\Omega~,~t\in [0,T]~,
\end{equation}
\item obtain the weight functions $y_{j}(\mathbf{x})$ from
$Q:=\Re(\mu(\mathbf{x},\omega))/\Im(\mu(\mathbf{x},\omega))$ via
\begin{equation}\label{3.35}
 \mu(\mathbf{x},\omega)=\mu_{R}(\mathbf{x})\left[ 1+
 \sum_{j=1}^{J}y_{j}(\mathbf{x})\frac{i\omega}{i\omega-\omega_{j}}\right]
~~;~~
 \mathbf{x}\in\Omega~,~\omega\in
 \left[ \frac{\omega_{max}}{100},\omega_{max}\right] ~,
\end{equation}
\item solve the coupled system of three first-order-in-time
partial differential equations
\begin{equation}\label{3.36}
 \mu_{R}(\mathbf{x})\nabla u(\mathbf{x},t)-
 \partial_{t}\mathbf{V}(\mathbf{x},t)=0~~;~~
 \mathbf{x}\in\Omega~,~t\in [0,T]~,
\end{equation}
\begin{equation}\label{3.37}
 \rho\partial_{t}u(\mathbf{x},t)-\nabla\cdot\mathbf{V}(\mathbf{x},t)=
 \rho g(\mathbf{x},t)+\sum_{j=1}\partial_{t}\eta_{j}(\mathbf{x},t)~~;~~
 \mathbf{x}\in\Omega~,~t\in [0,T]~,
\end{equation}
\begin{equation}\label{3.38}
\partial_{t}\eta_{j}(\mathbf{x},t)+ \omega_{j}\eta_{j}(\mathbf{x},t)=
\frac{1}{\rho(\mathbf{x})}
 \nabla\cdot\left ( y_{j}(\mathbf{x})\mathbf{V}(\mathbf{x},t)\right)
 ~~;~~
 \mathbf{x}\in\Omega~,~t\in [0,T]~,
\end{equation}
for~ the ~three ~unknown  ~~functions  ~~$u(\mathbf{x},t)$,~
$\mathbf{V}(\mathbf{x},t)$, ~ and~~ $\boldsymbol{\eta}:=
\{\eta_{j}(\mathbf{x},t)~;~j=1,2,..,J\}$.
\end{itemize}
\subsection{Description of the cities and methods of analysis}\label{s22}
The densities in the bedrock, soft layer and blocks (together with their foundations) where chosen to be: $2000kg.m^{-3}$, $1300kg.m^{-3}$, and $325kg.m^{-3}$ respectively. The bulk shear wave velocities in these three media were taken to be $600m.s^{-1}$, $60m.s^{-1}$, and $100m.s^{-1}$ (instead of $200m.s^{-1}$ in \cite{Wir2003}) respectively, and the quality factors were chosen to be $+\infty$, $30$, and $100$ respectively. The foundation depth of the blocks was $10m$ and the soft layer thickness was taken to be $50m$. The block widths and heights ranged over $30-60m$ and $50-70m$  respectively (see Table \ref{t1}). The block separations ranged over $60-100m$ for $C^{1}$, $30-50m$ for $C^{2}$ and $10-30m$ for $C^{3}$ (see Table \ref{t1}). 

We computed the seismic response both in the absence and presence of the blocks. The media below  ground level were the same in both of these cases. Most of the aforementioned parameters are close to those of \cite{Wir1996} and \cite{Wir2003}, and are fairly representative of the blocks and substratum at downtown sites in Mexico City. The computational domain was a $3500m \times 3500m$  square discretized by a grid of 700 nodes in each dimension. This domain was surrounded by a PML layer \cite{Col2001} 29 nodes thick, and 943 nodes where placed on the free surface \cite{Bec2001} .

To give a measure of the vulnerability of the blocks of the city, we computed the vulnerability index $R_{j}$, introduced in \cite{Wir2003}:
\begin{equation}
R_{j}=\frac{\int_{0}^{T} \left| \partial_{t} u(\mathbf{x}_{j},t) \right|^{2} }{\int_{0}^{T} \left| \partial_{t} u^{0} (\mathbf{x}_{0},t) \right|^{2} }
\end{equation}
wherein: $T$ is the time interval of significant shaking (taken herein to be $200s$), $\left| \partial_{t}u(\mathbf{x}_{j},t)\right|^{2}$ the modulus-squared particle velocity at the midpoint of the $j$-th block, and $\left| \partial_{t}u^{0}(\mathbf{x}_{0},t)\right|^{2}$  the same quantity measured on the ground in the absence of all the blocks.

To give a measure of the strength of ground shaking, we computed the index $R_{j,j+1}$:
\begin{equation}
R_{j}=\frac{\int_{0}^{T} \left| \partial_{t}u(\mathbf{x}_{j,j+1},t)\right|^{2}}{\int_{0}^{T} \left| \partial_{t}u^{0}(\mathbf{x}_{0},t)\right|^{2}}
\end{equation}
wherein: $T$ is as previously, $\left| \partial_{t}u(\mathbf{x}_{j,j+1},t)\right|^{2}$ is the modulus-squared particle velocity at the center of the ground segment between the the $j-th$ block and the $j+1-th$ block, and $\left| \partial_{t}u^{0}(\mathbf{x}_{0},t)\right|^{2}$ is as previously.
\subsection{Validation of the method}
\begin{figure}[H]
\includegraphics[width=7cm]{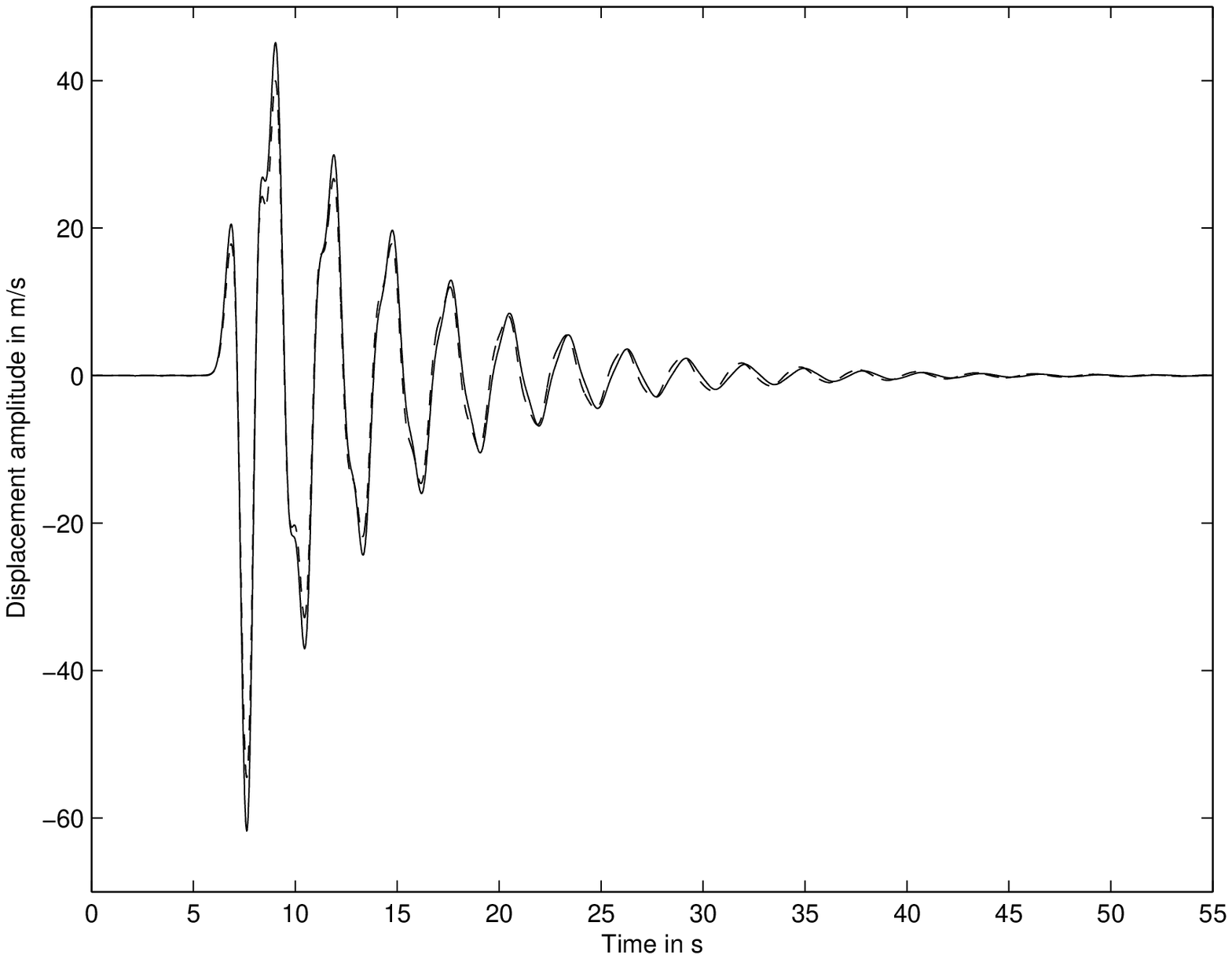}\hfill
\includegraphics[width=7cm]{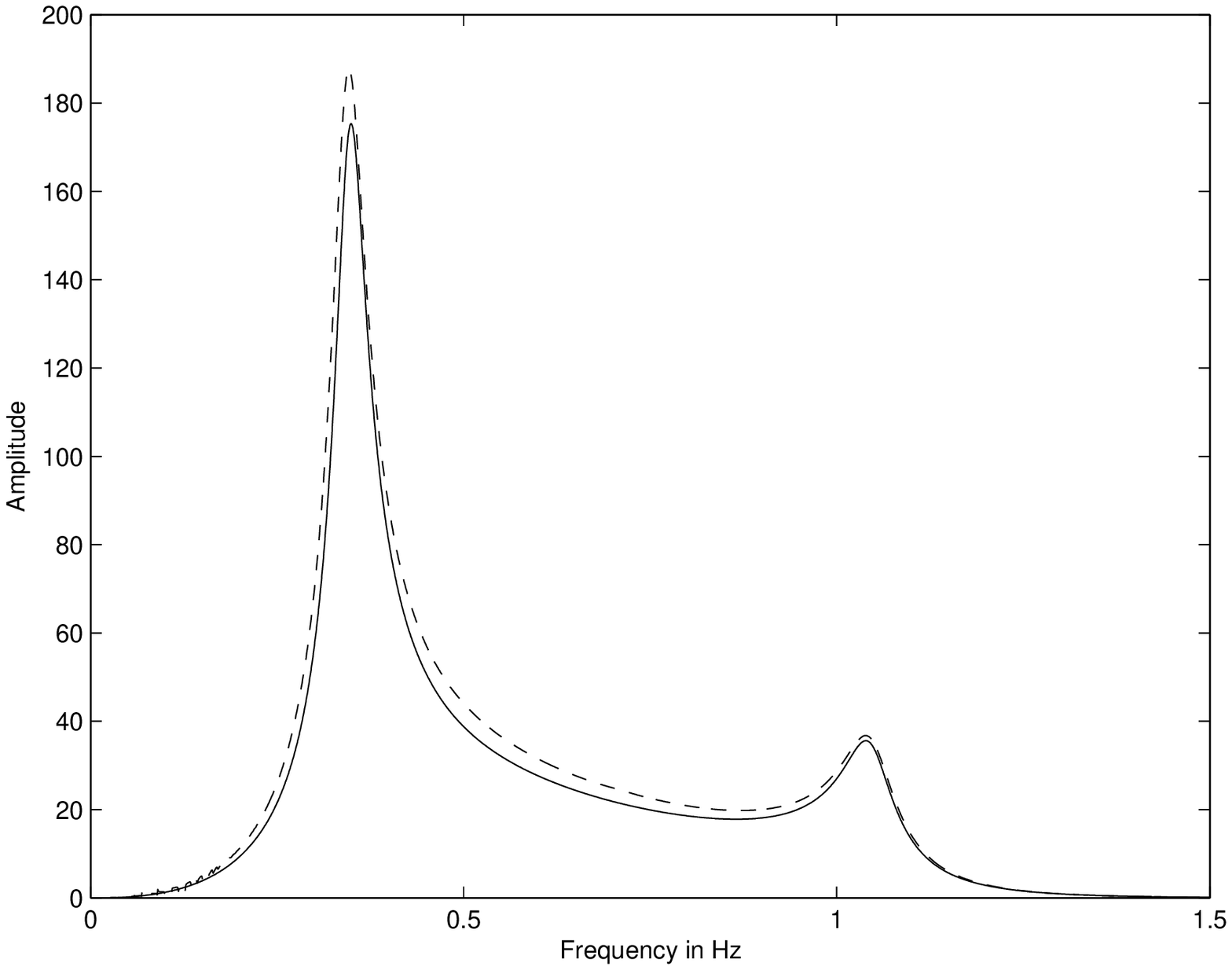}
\caption{Time history of the displacement (left panel) and displacement spectrum (right panel). Solid-line: numerical solution, dashed-line: semi-analytical solution}
\label{valid}
\end{figure}

We validated the time domain method on the canonical example of flat ground with no blocks. The viscoelastic modulus of the soft layer is given by Kjartansson's formula \cite{Kjar1979}:
\begin{equation}
\mu_{1}(\mathbf{x},\omega)= \left|\mu_{R} (\mathbf{x},\omega_{R}) \right|\left(\frac{-i\omega}{\omega_{R}} \right)^{\frac{2}{\pi} \arctan \left( Q_{1}^{-1} \right)}
\label{Kjar}
\end{equation}
The source position was $\mathbf{x}_{S}=(0m,3000m)$, and we computed the displacement at the point just below the hypothetic building $B_{1}$.
In Fig.\ref{valid}, the small differences between the semi-analytical \cite{grwi04} and numerical solution are due to: i) the fact that the semi-analytical solution involves numerical integration, and ii) discretization errors in our time-domain method.
\section{Results}
The aim of this work was first to prove that amplification, beatings, long duration, and spatial variability of response are observed even when the  media (in the layer and blocks) are dissipative, and that essentially the same phenomena are involved in both the dissipative and non-dissipative cases.

\begin{figure}[H]
\begin{center}
\includegraphics[width=8cm]{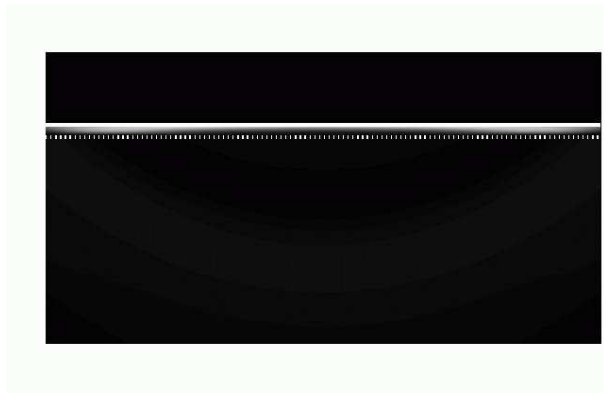}
\caption{Snapshot of the total (i.e., incident plus scattered) displacement field, at $t=15s$, for  flat ground underlain by a soft layer and hard half space.}
\end{center}
\label{Snaplayer}
\end{figure}

The snapshot of the modulus of the total displacement field in  Fig. \ref{Snaplayer} pertains to the case in which there are no blocks in the city, for the dissipative layer case. As mentioned in \cite{Wir1996}, \cite{Wir2003}, 
\cite{grwi04}, one does not expect the Love modes to be excited to any great extent when the source is far from the layer. This is what is actually observed in Fig. \ref{Snaplayer} since the displacement field in the soft layer is not typical of that of a Love mode.

\begin{figure}[H]
\begin{center}
\includegraphics[width=12cm]{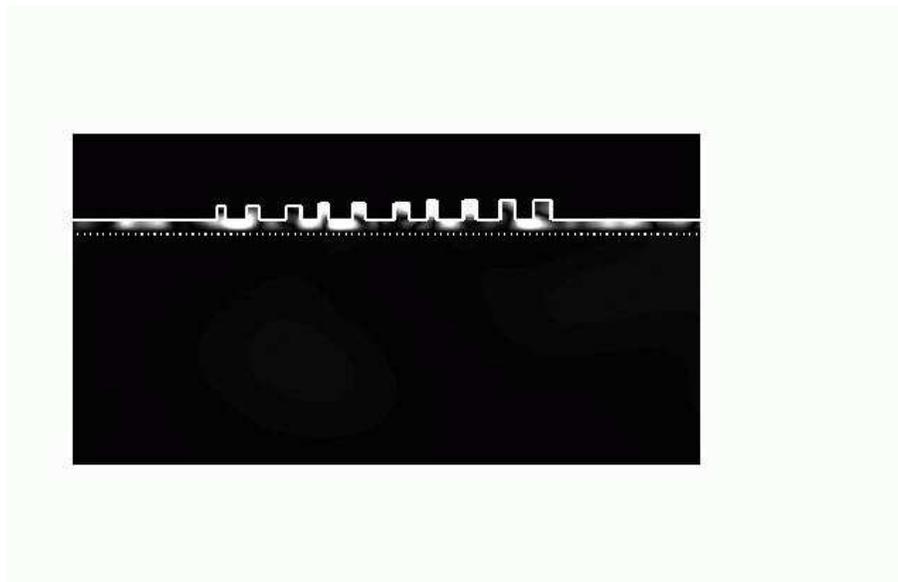}
\caption{Snapshot of the total (i.e., incident plus scattered) displacement field, at $t=15s$, for  city $C^{1}$ with 10 blocks.}
\end{center}
\label{Snapblock}
\end{figure}

Fig. \ref{Snapblock} depicts the snapshot for a city overlying a dissipative layer with ten blocks. One observes a series of hot spots inside the low-velocity layer, constituting an indication of something like a standing wave in the layer betraying the excitation of a quasi-Love mode \cite{wiko93}, \cite{sedu00}, \cite{grwi04}. This configuration also gives rise to rather large response  in the blocks and on the ground, as was previously observed in \cite{Wir2003} for non-dissipative media, and in \cite{Wir1996} for a periodic distribution of identical blocks.

\begin{figure}[H]
\begin{center}
\includegraphics[width=12cm]{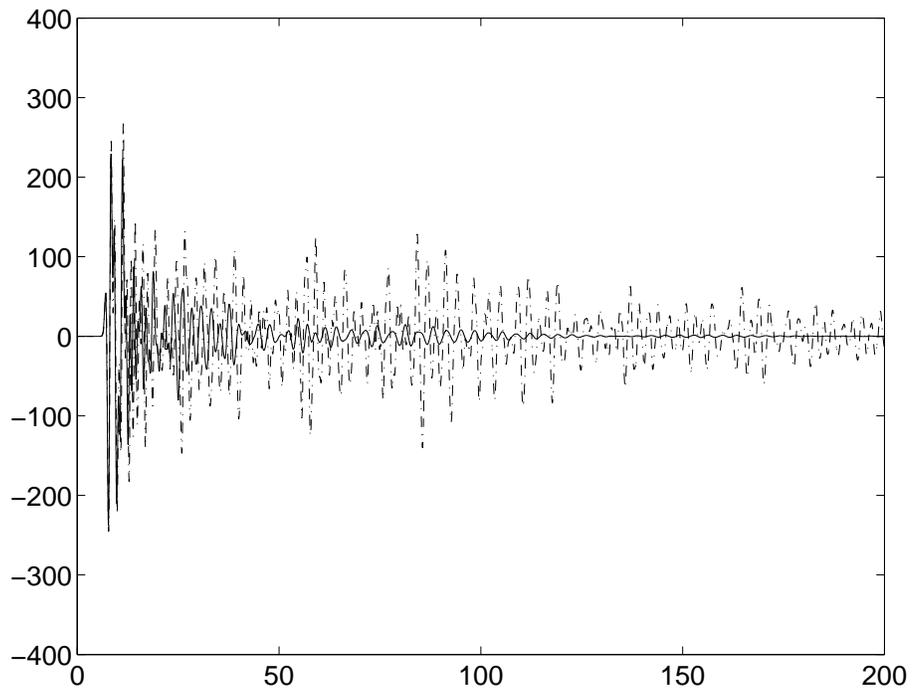}
\caption{Time record of the velocity on top of  building no. 6, for  city  $C^{1}$ with 10 blocks. Solid line: dissipative media, dashed line: non-dissipative media.}
\end{center}
\label{CompV6}
\end{figure}

\begin{figure}[H]
\includegraphics[width=4.5cm]{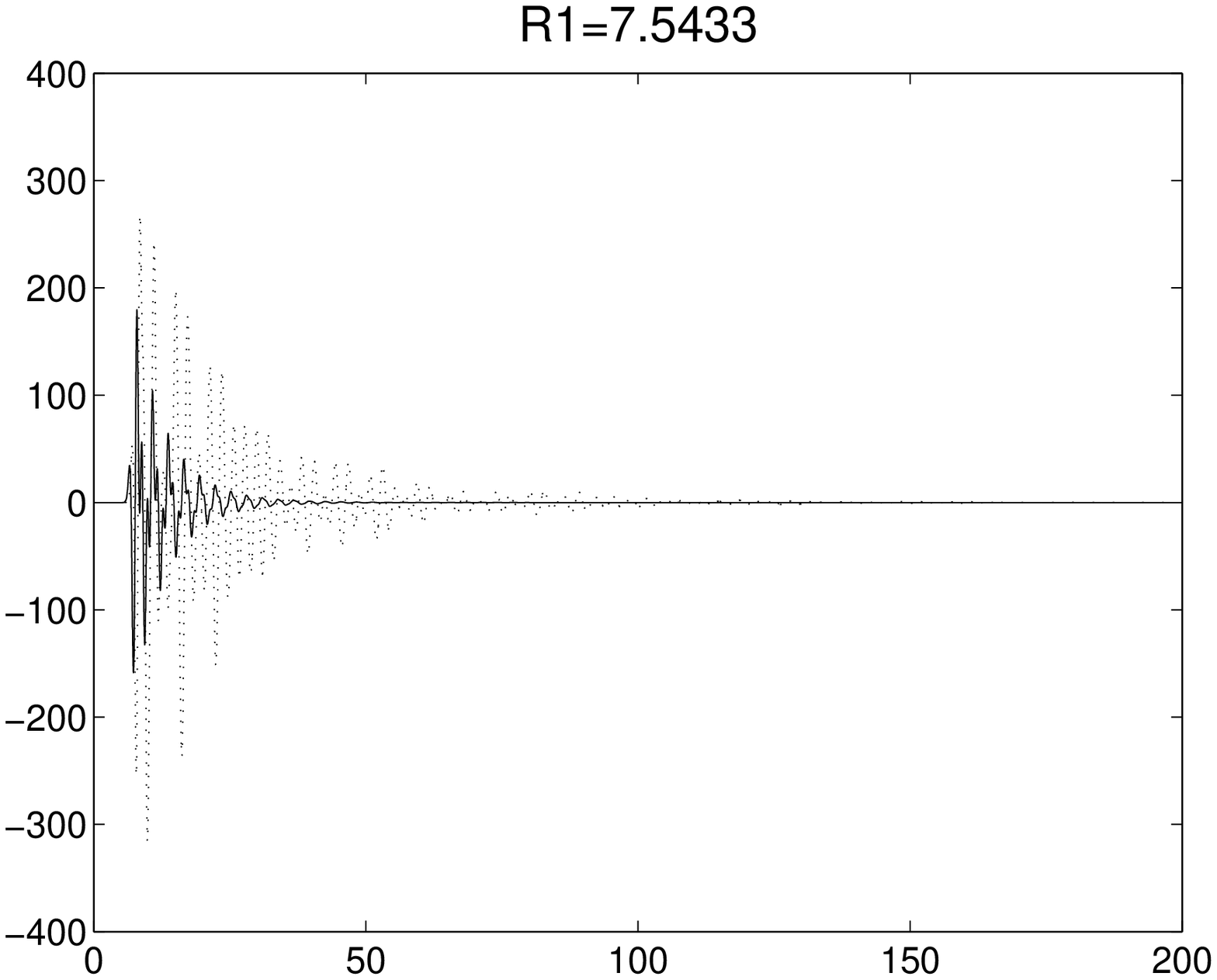}\hfill
\includegraphics[width=4.5cm]{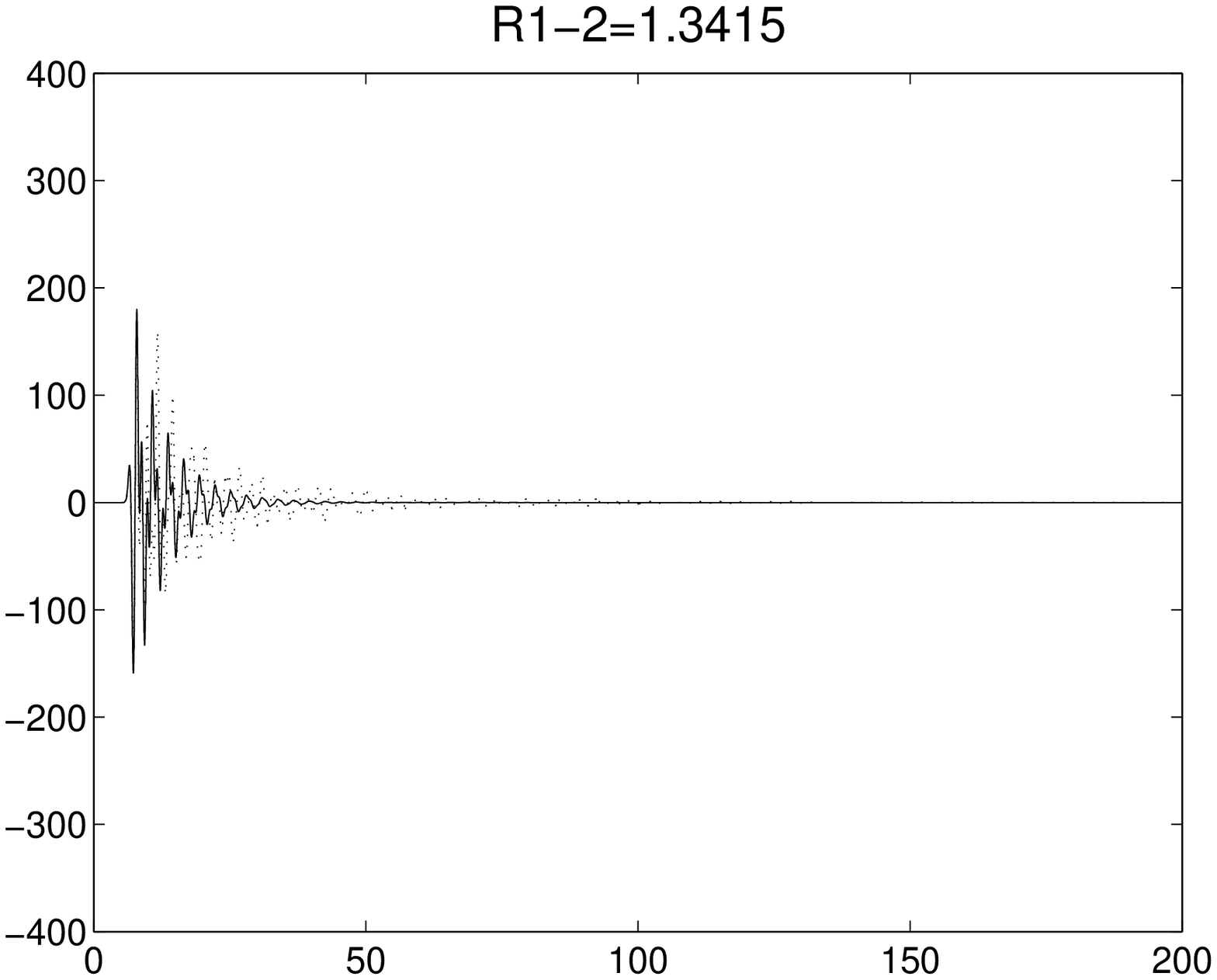}\hfill
\includegraphics[width=4.5cm]{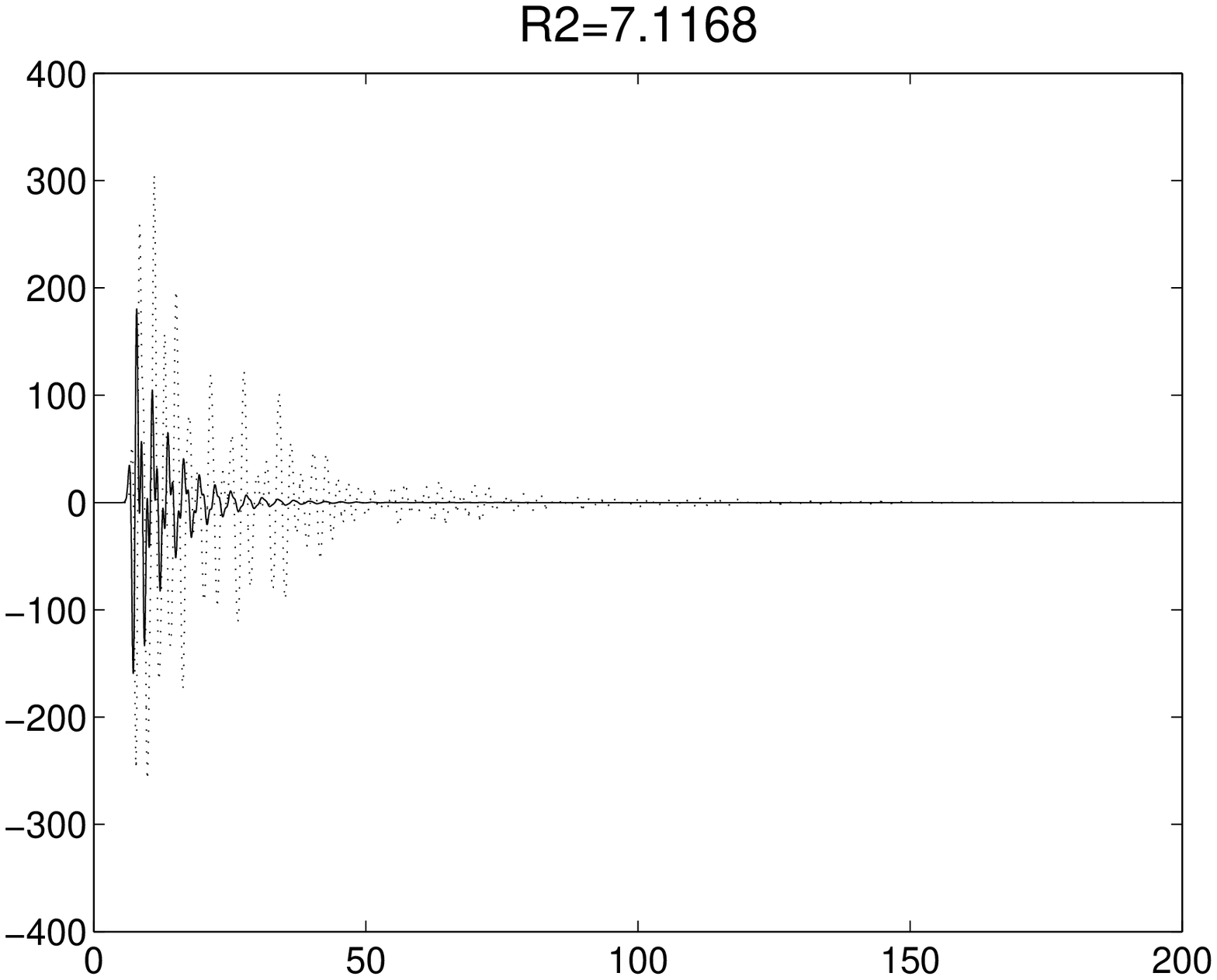}
\includegraphics[width=4.5cm]{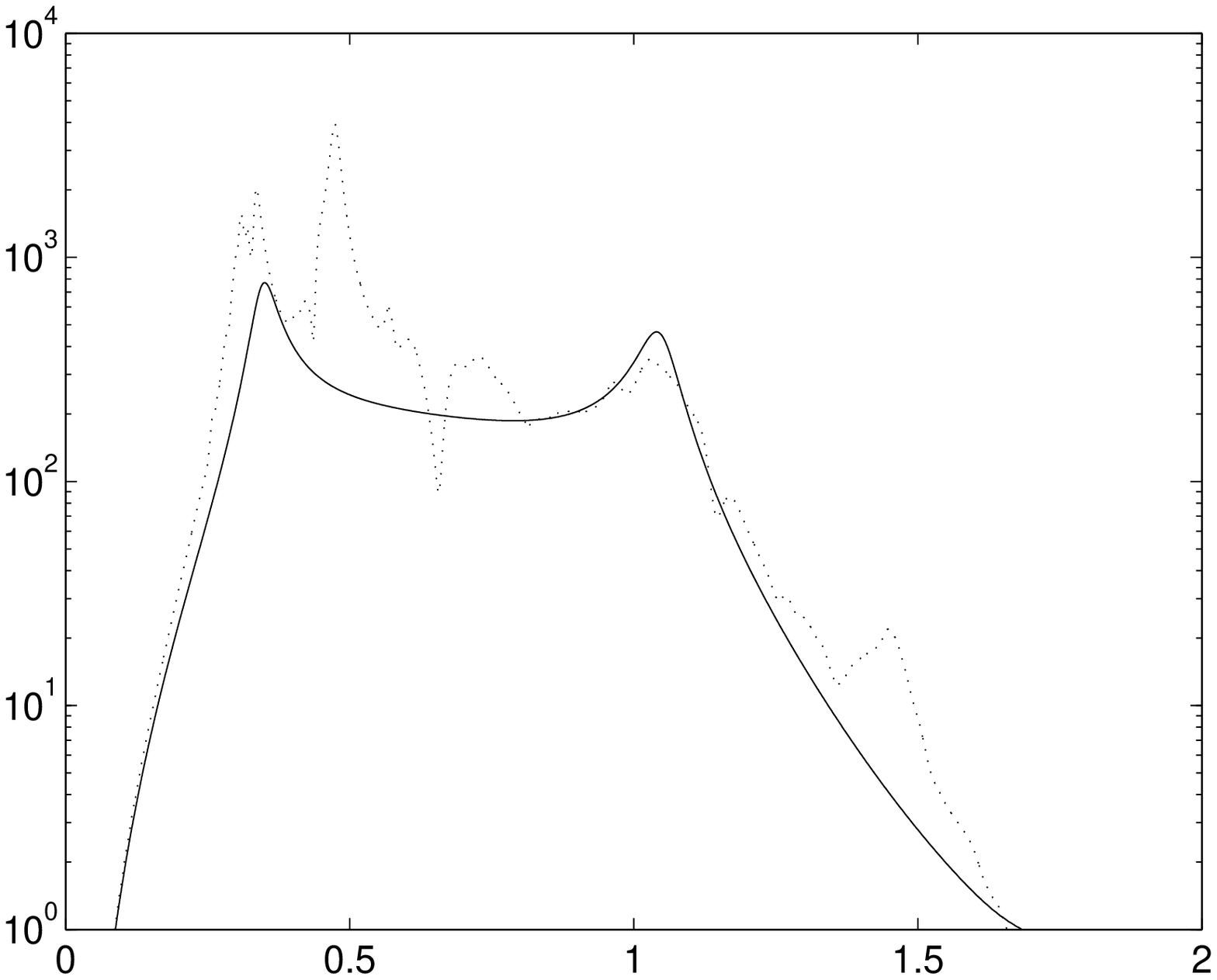}\hfill
\includegraphics[width=4.5cm]{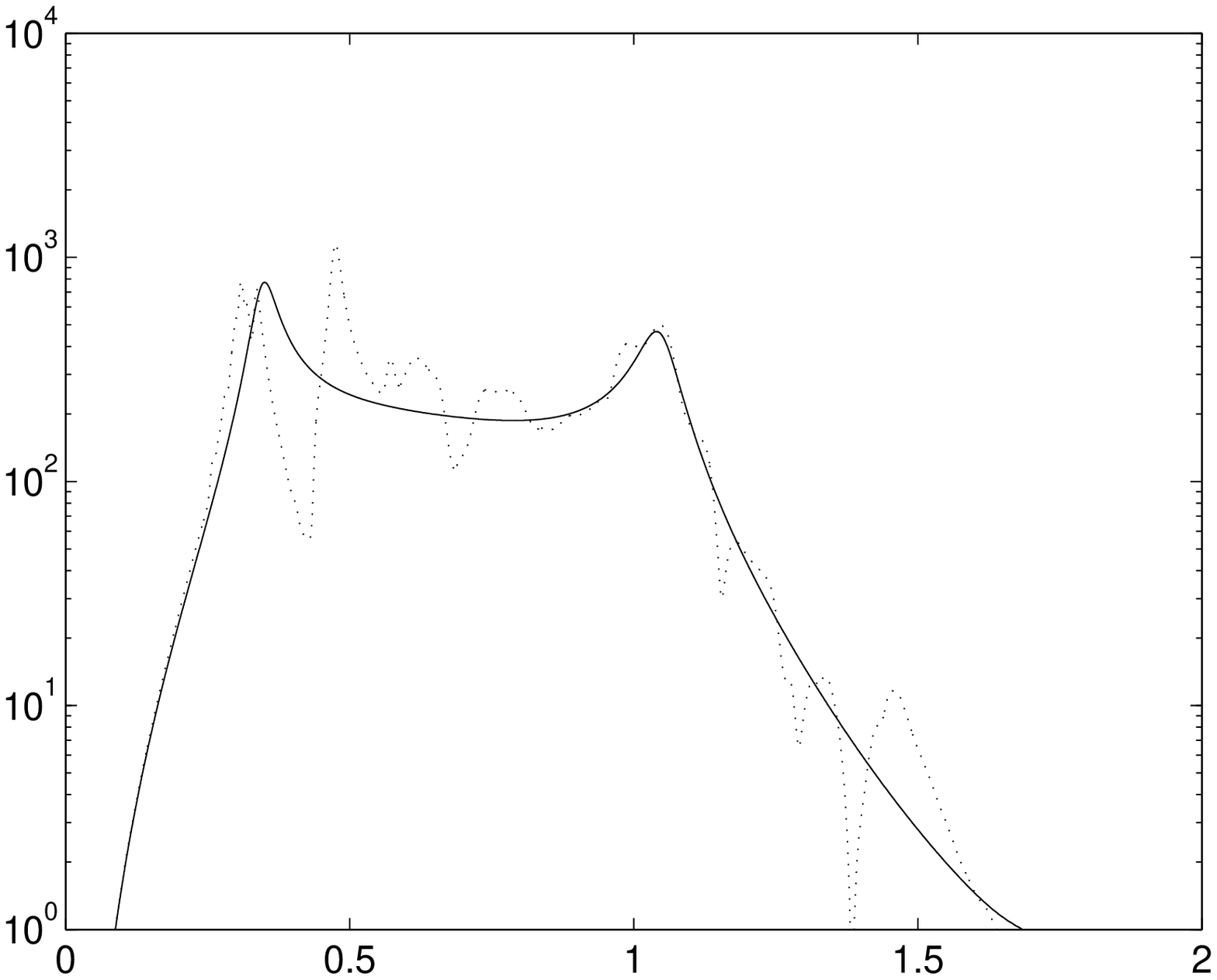}\hfill
\includegraphics[width=4.5cm]{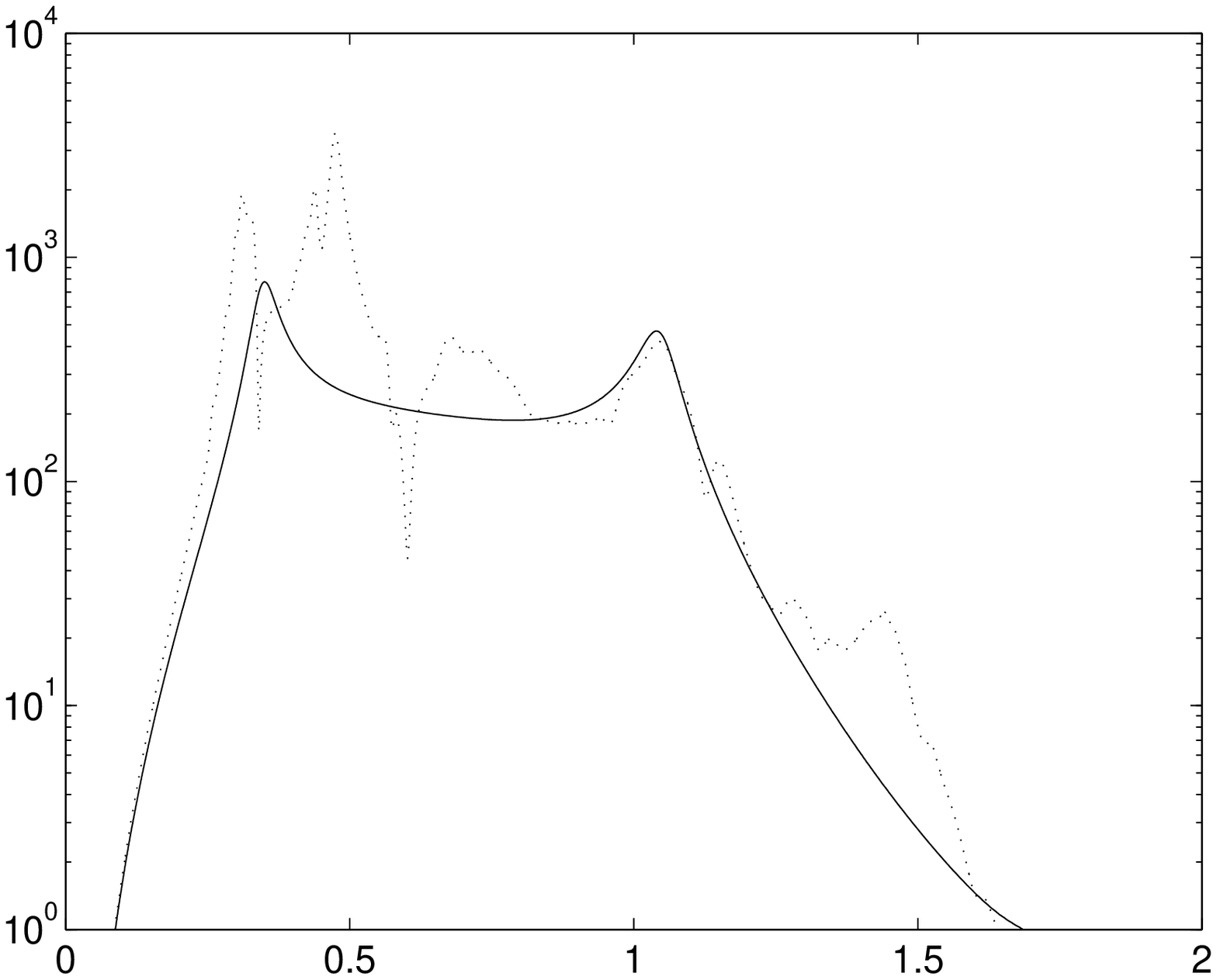}
\includegraphics[width=4.5cm]{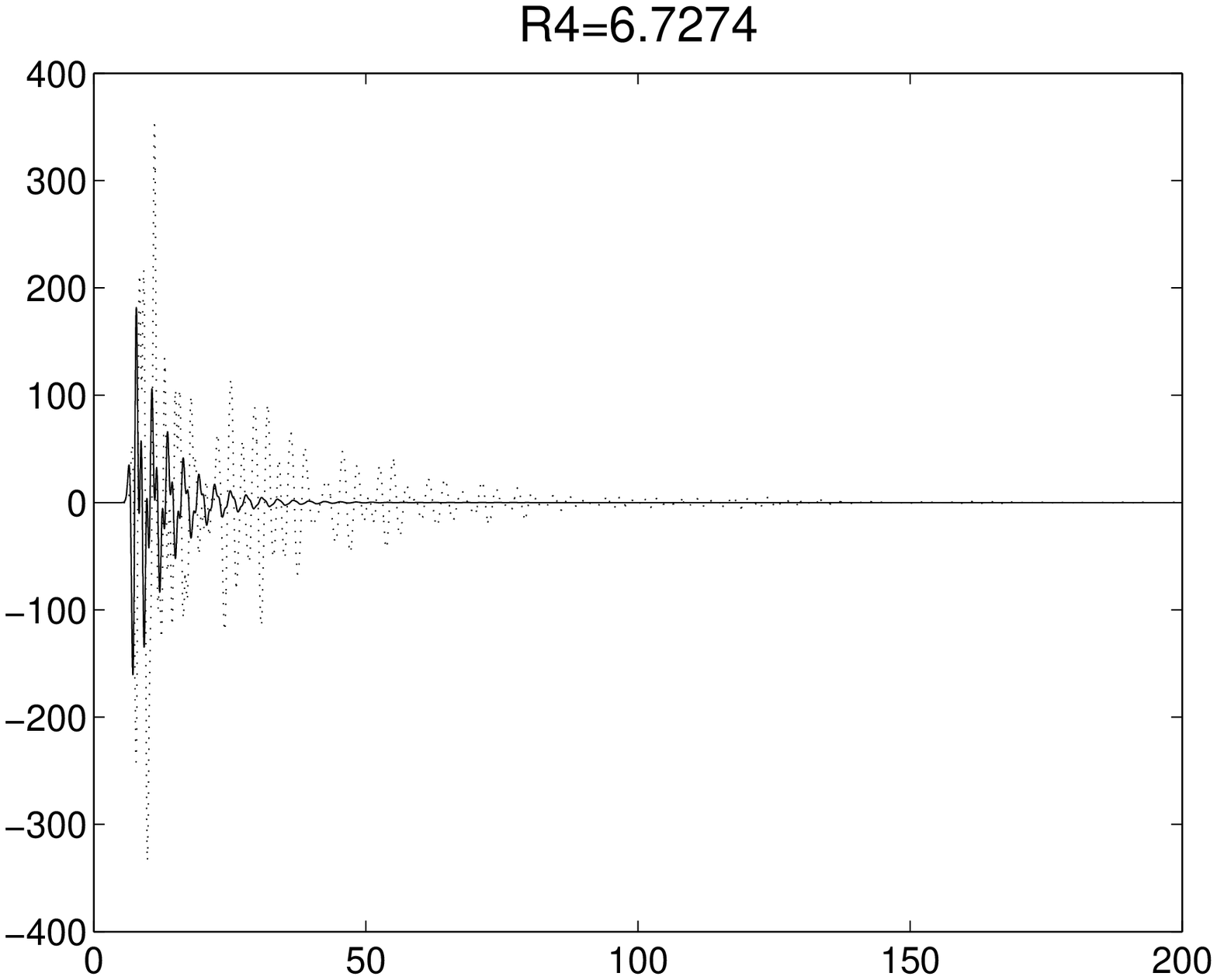}\hfill
\includegraphics[width=4.5cm]{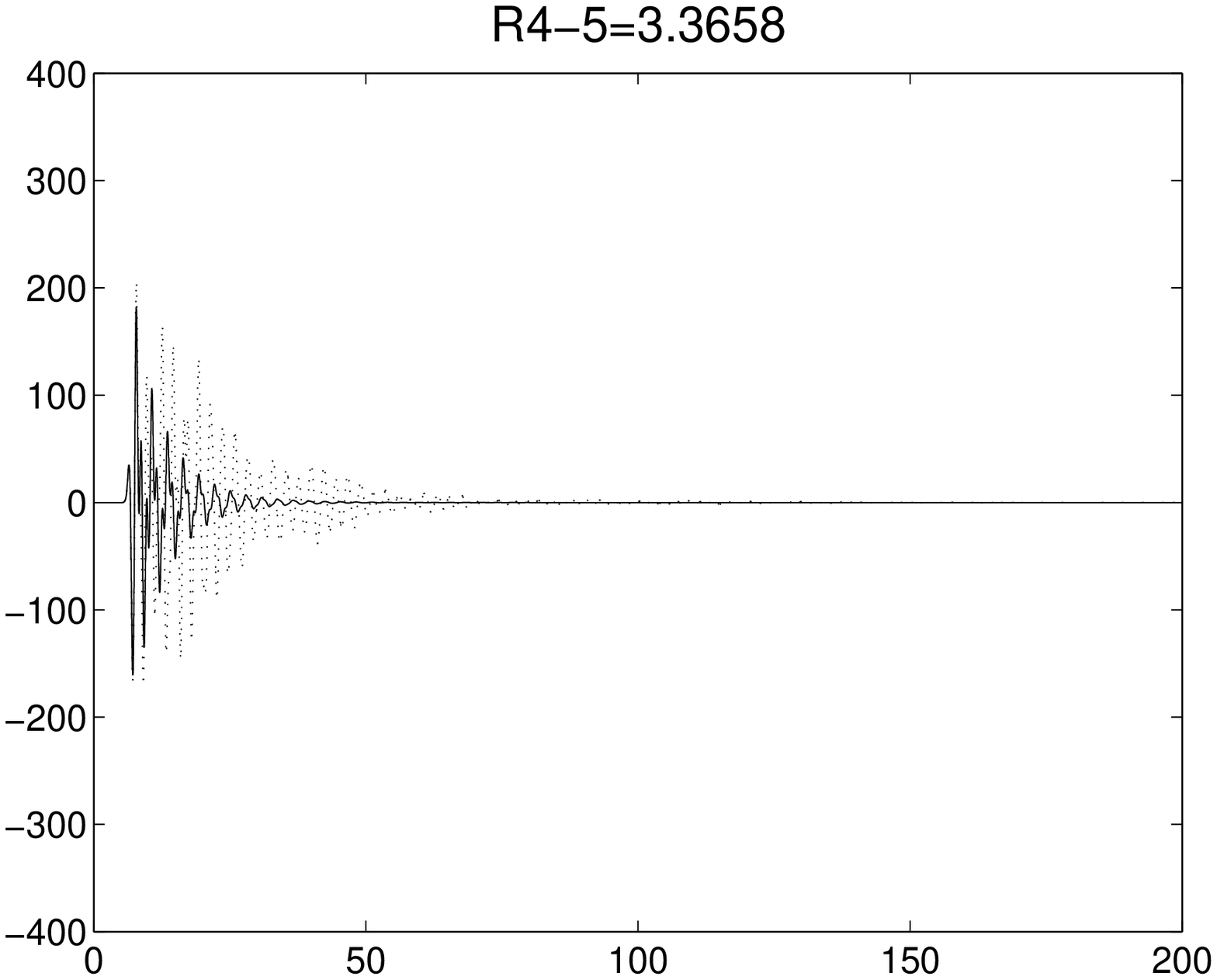}\hfill
\includegraphics[width=4.5cm]{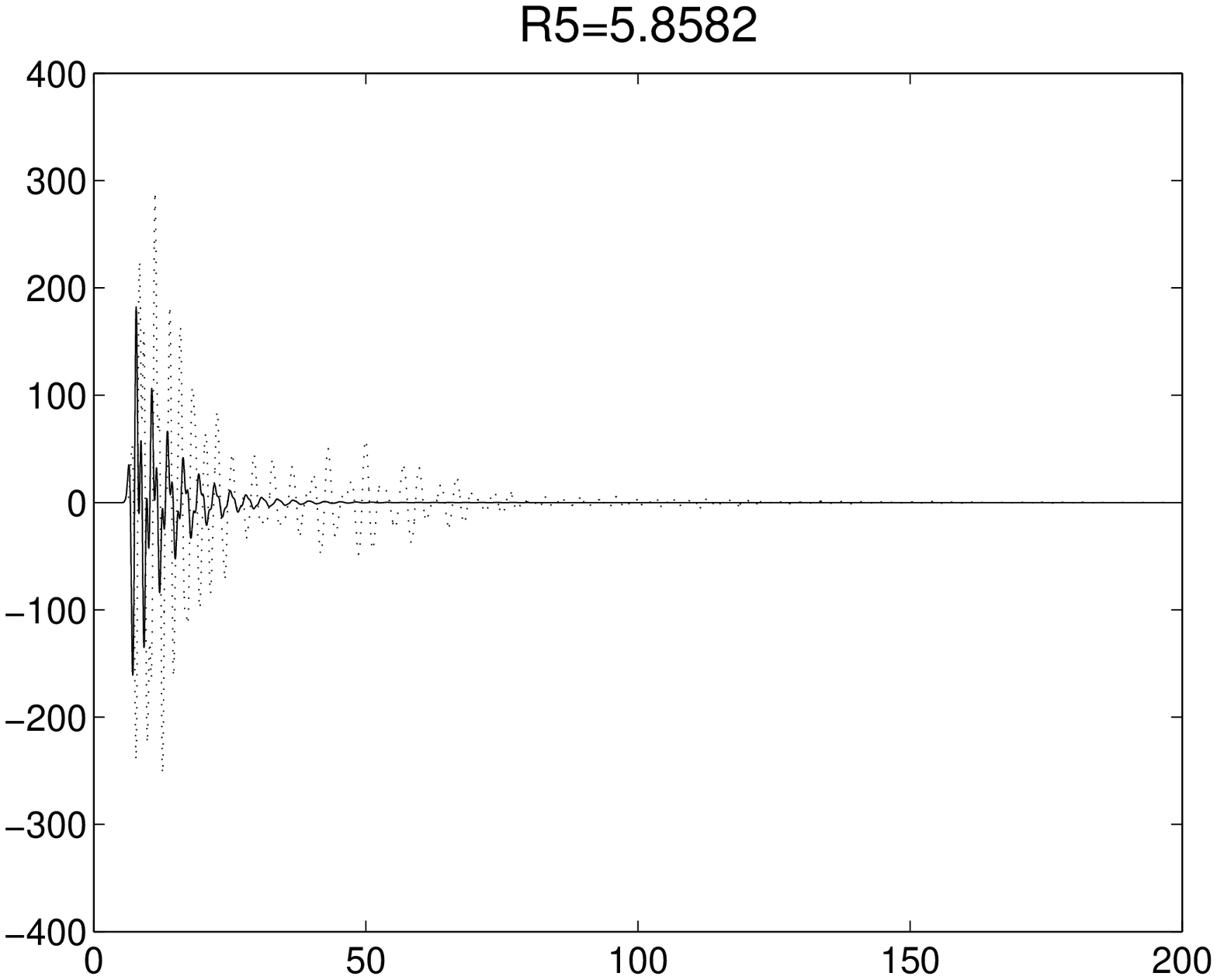}\hfill
\includegraphics[width=4.5cm]{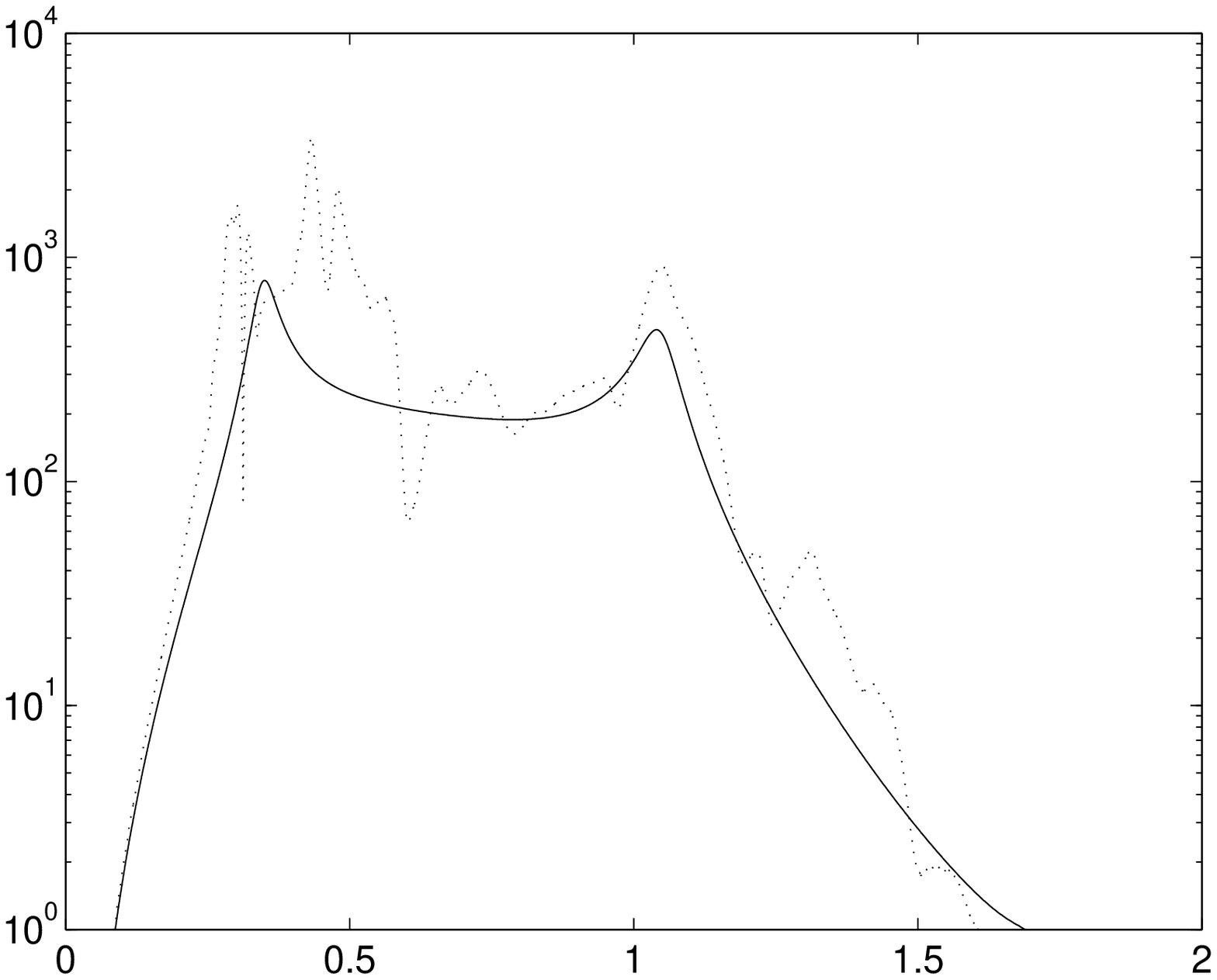}\hfill
\includegraphics[width=4.5cm]{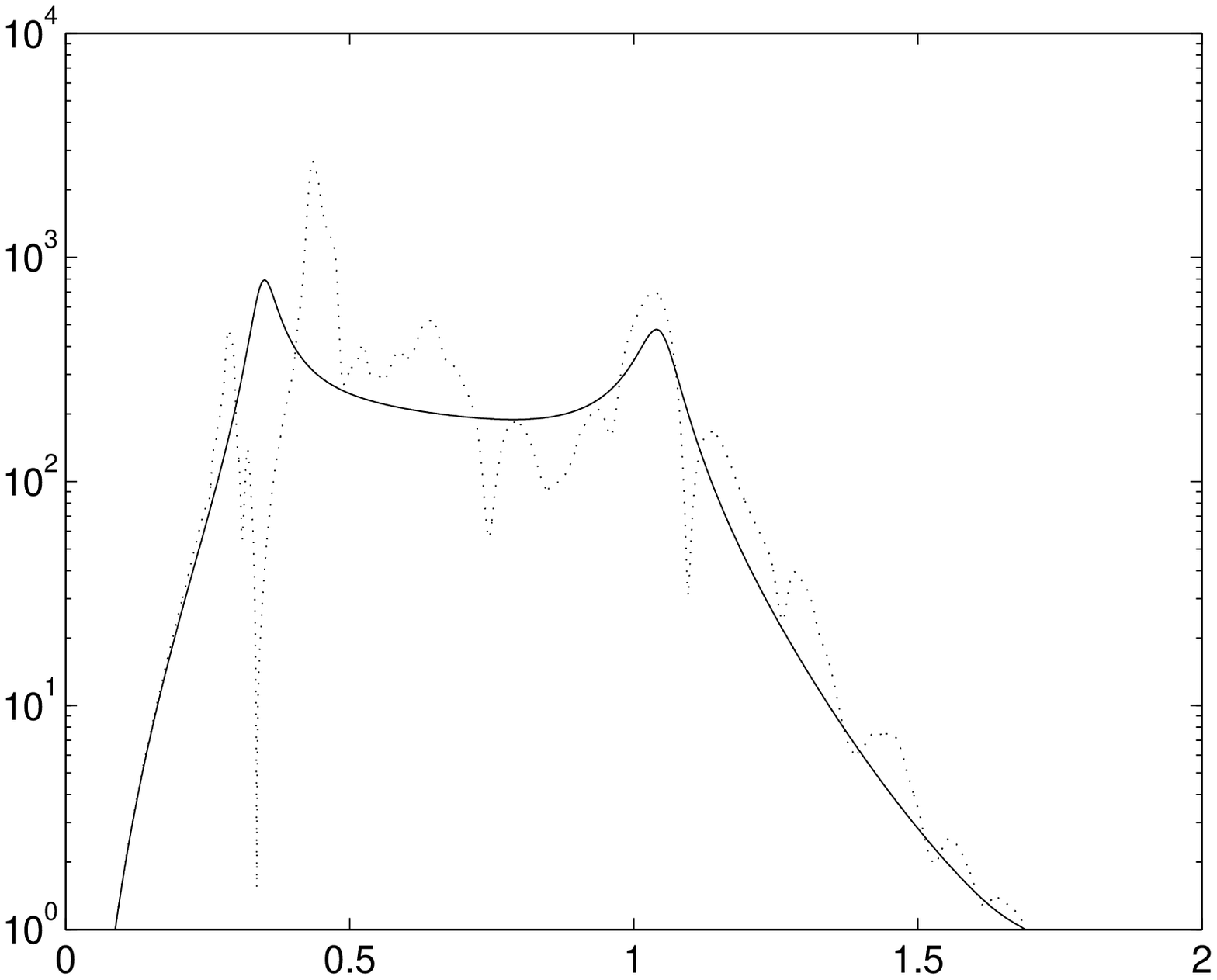}\hfill
\includegraphics[width=4.5cm]{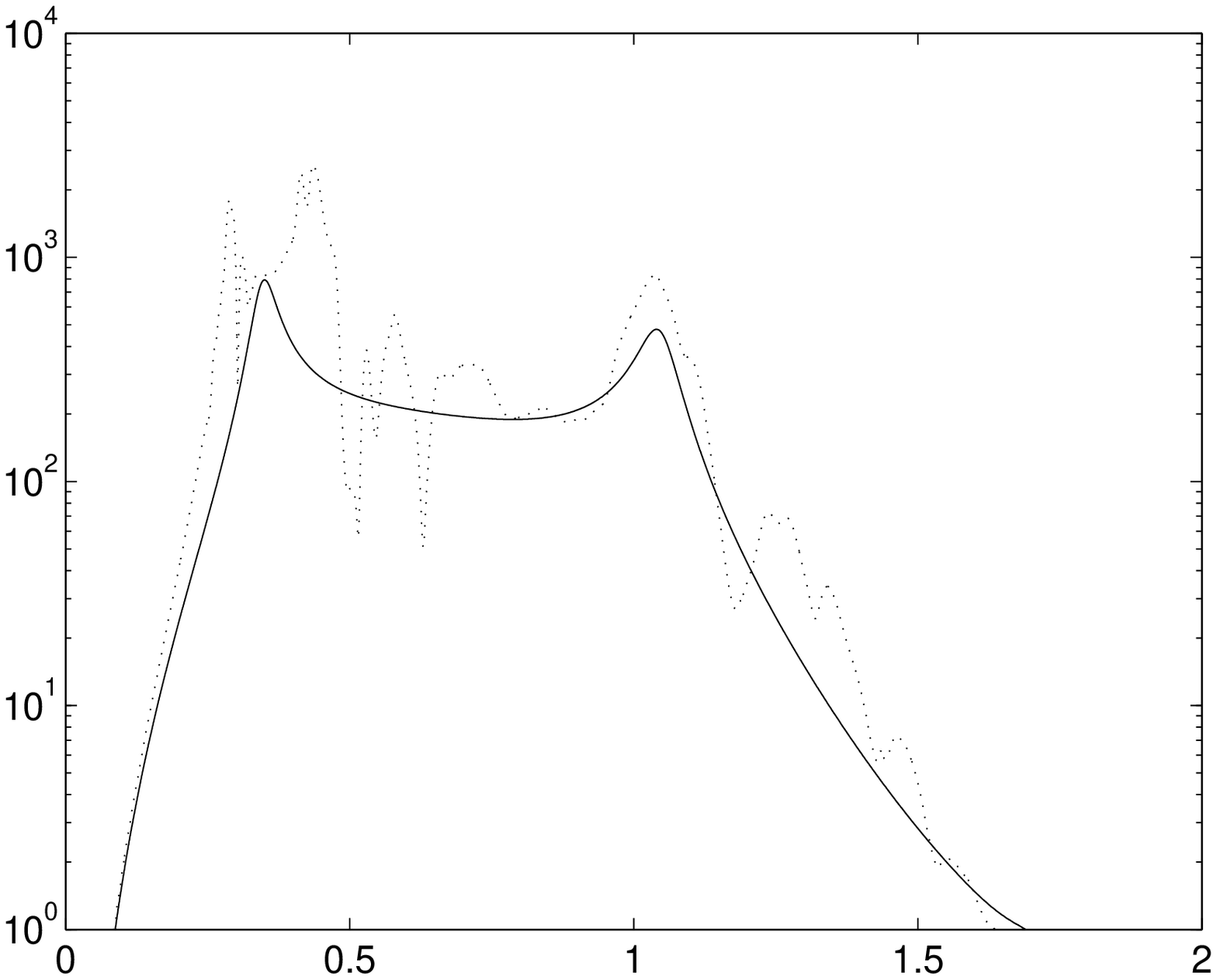}\hfill
\caption{Time records of total particle velocity for  `city' $C^{1}$ with ten blocks having different spacing $d_{j,j+1}^{1}$ ($j=1,2,...,9$). Each row of the figure depicts the particle velocity (in $s$): at the center of the top of the $j$-th block (left), the center of the ground segment between the $j$-th and the $j+1$-th block (middle) and at the top of the $j+1$-th block (right). The solid curves in all the subfigures represent the particle velocity at ground level in the absence of the blocks. The vulnerability indices $R_{j}$ at the top of the $j$-th block and $R_{j,j+1}$  on the ground between the $j$-th and the $j+1$-th block, are indicated at the top of each subfigure. The abscissas designate time, and range from $0$ to $200s$. Note that the scales of the ordinates do not vary from one subfigure to another. Below each time record appears the modulus of the velocity spectrum of the previous quantity. The abscissas designate frequency, and range from $0$ to $2Hz$.}
\label{TimeC1}
\end{figure}

\begin{figure}[H]
\includegraphics[width=4.5cm]{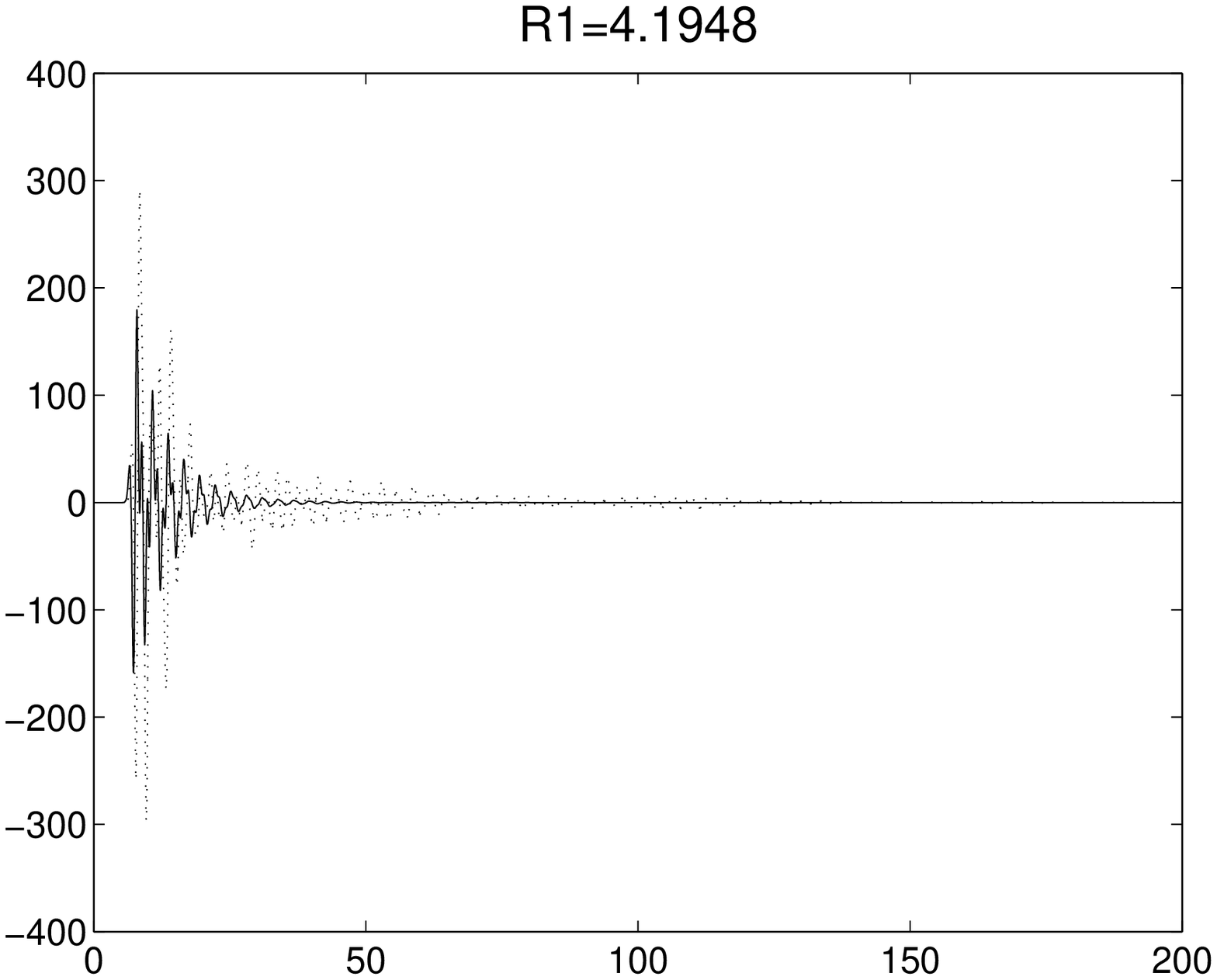}\hfill
\includegraphics[width=4.5cm]{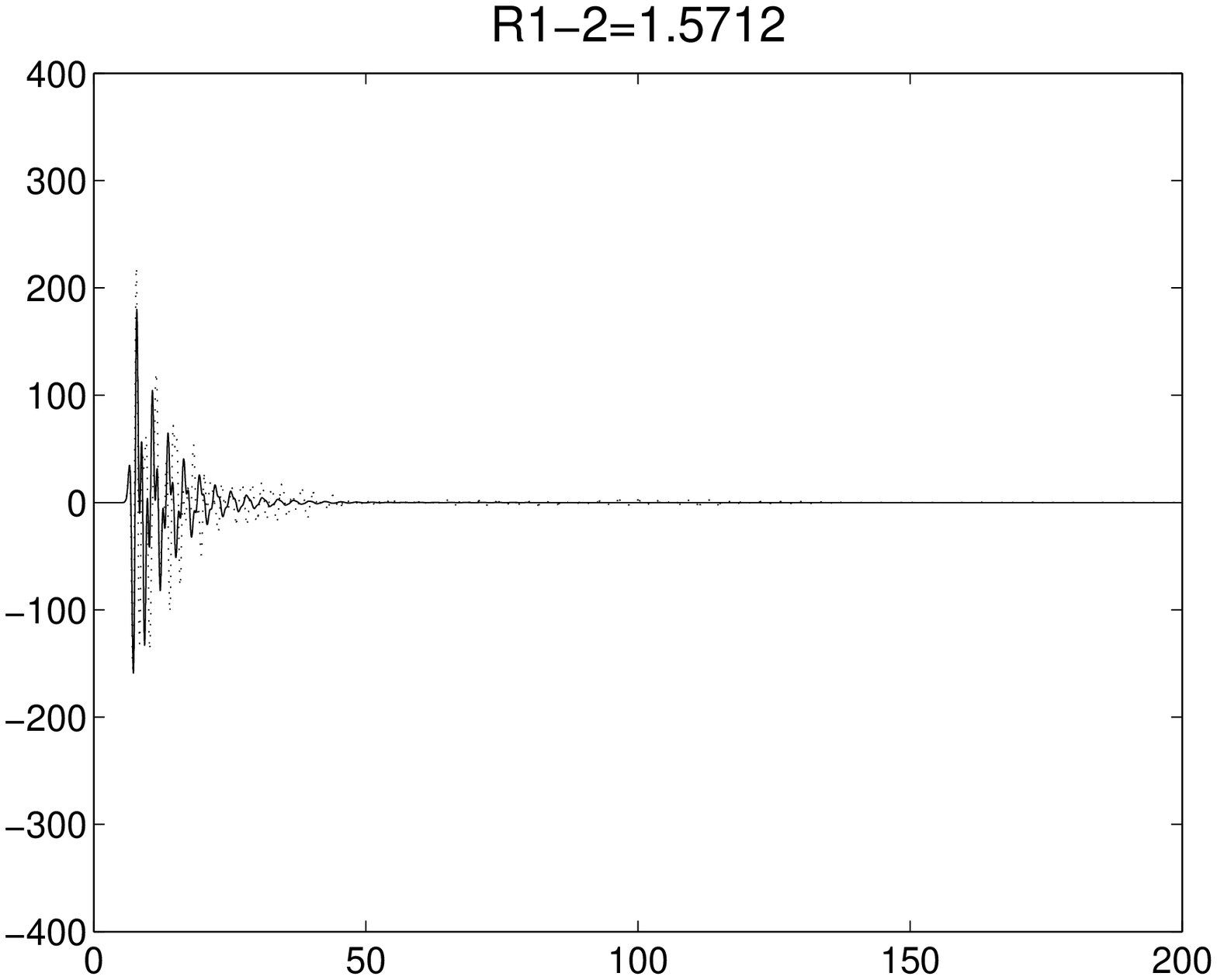}\hfill
\includegraphics[width=4.5cm]{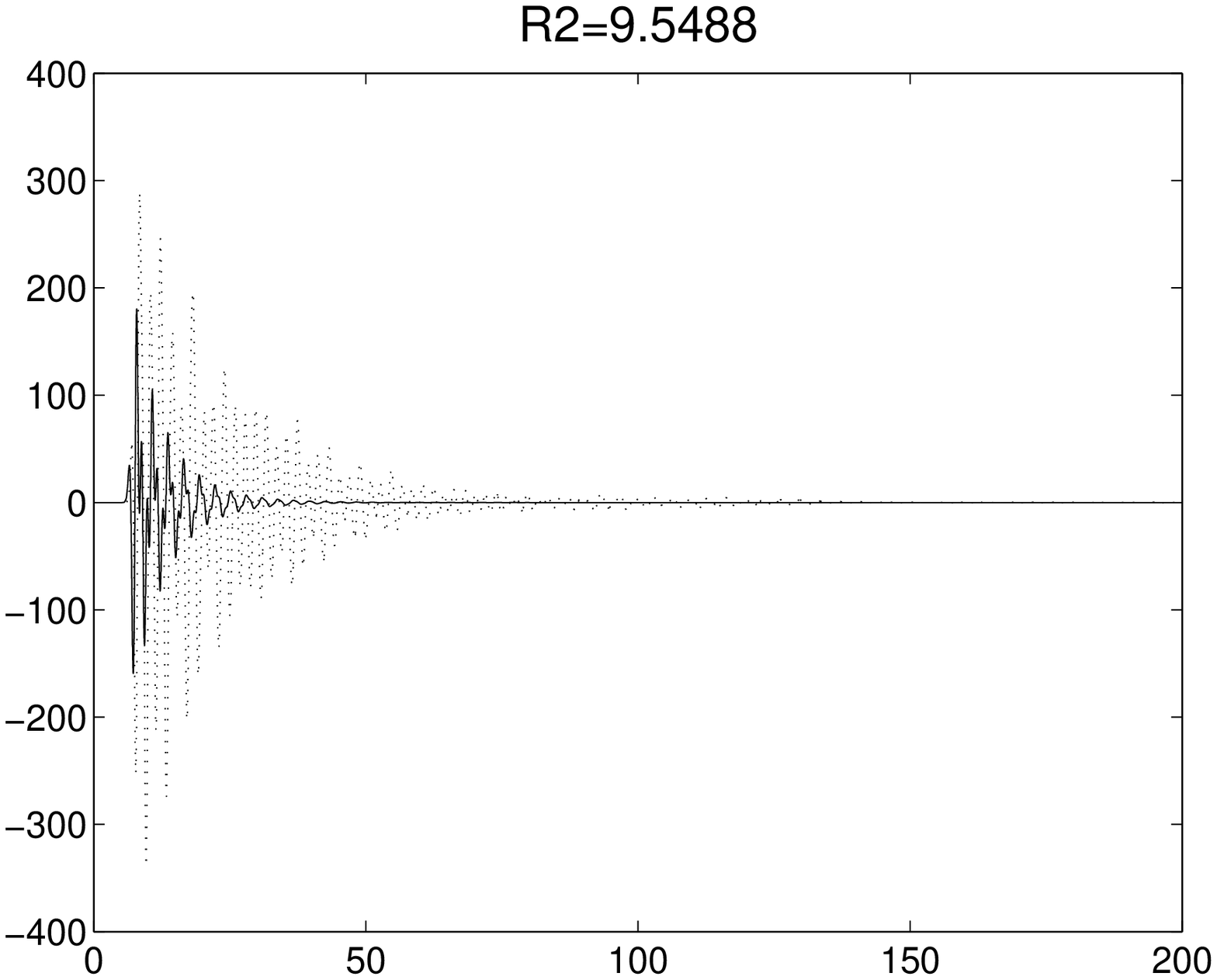}
\includegraphics[width=4.5cm]{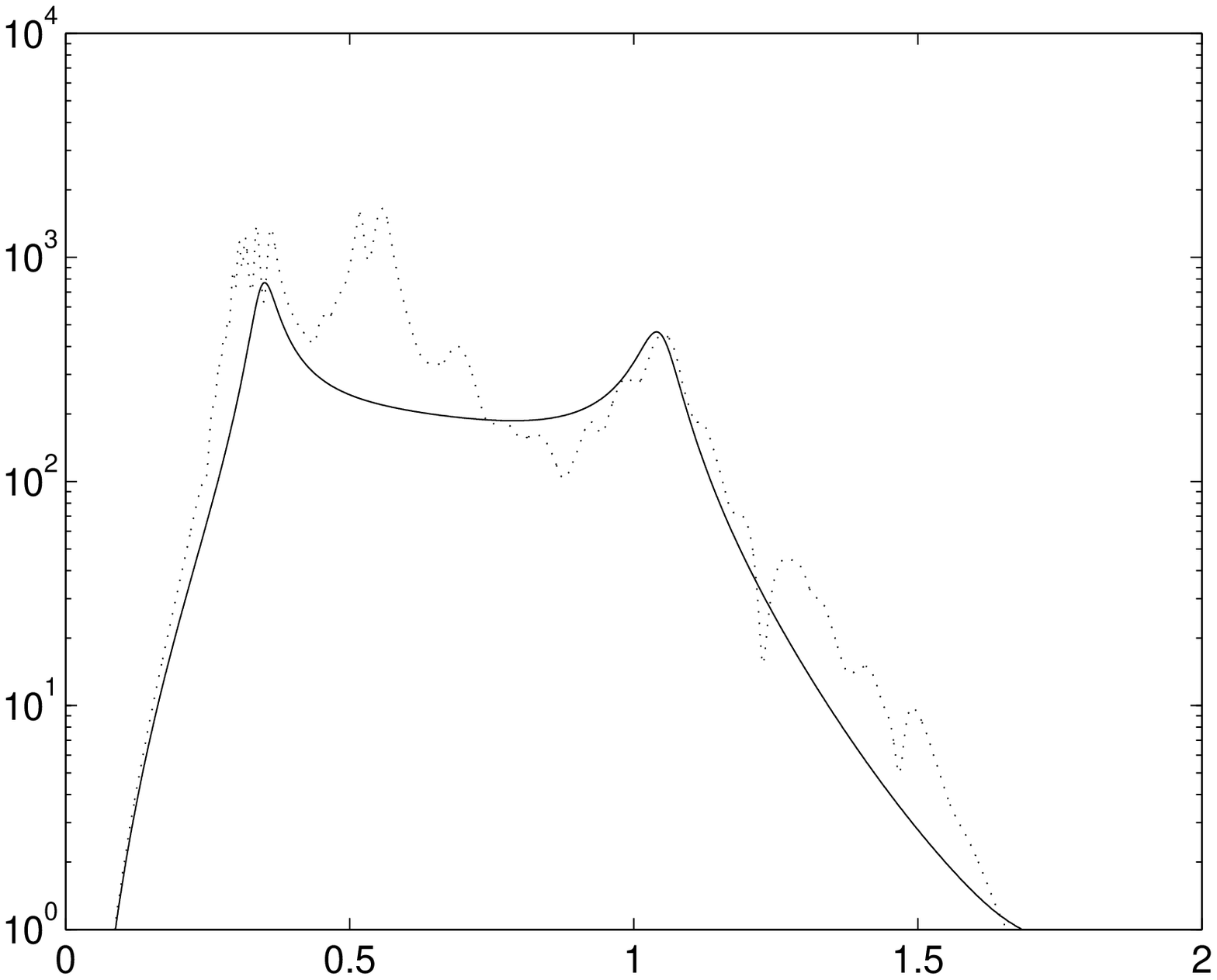}\hfill
\includegraphics[width=4.5cm]{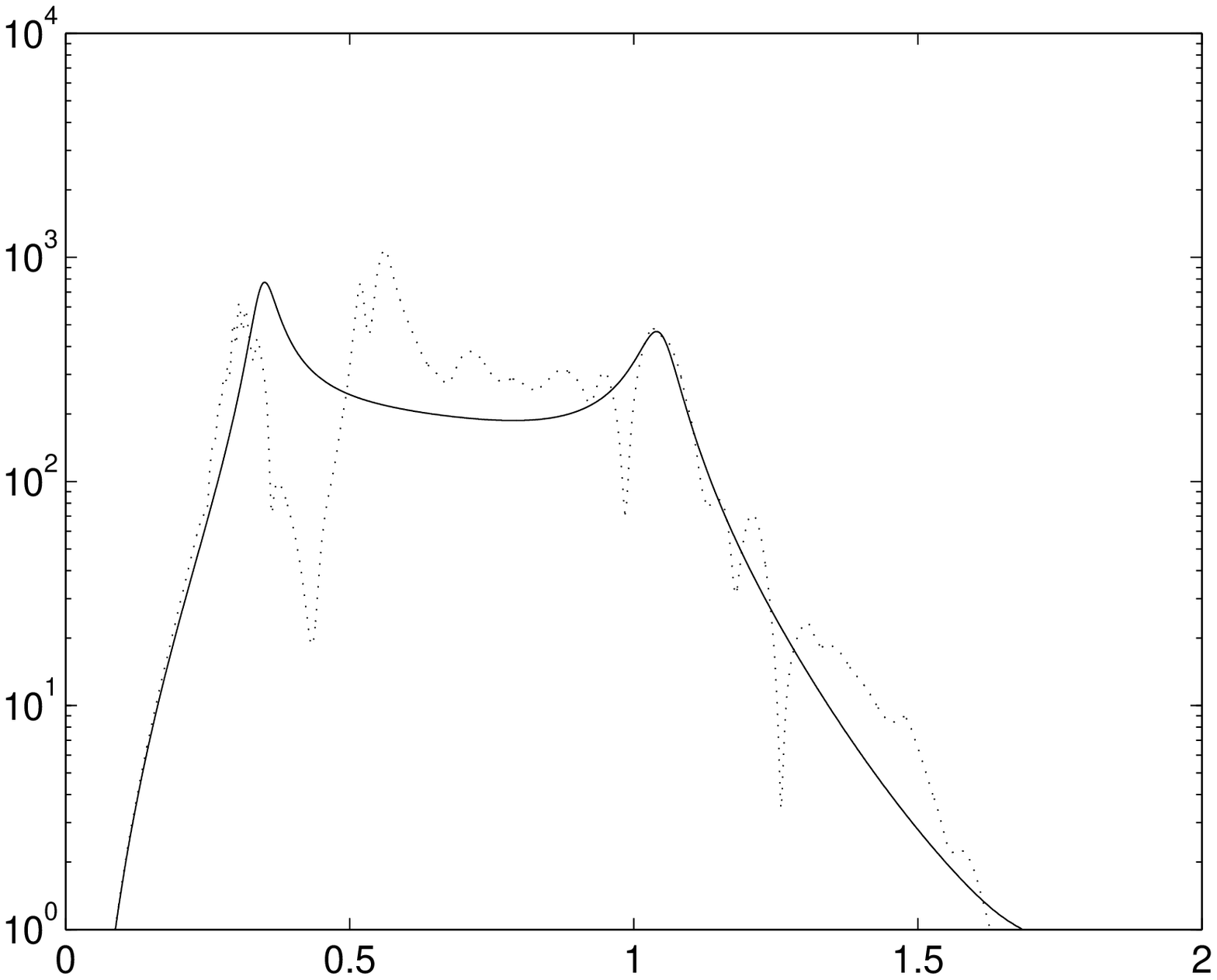}\hfill
\includegraphics[width=4.5cm]{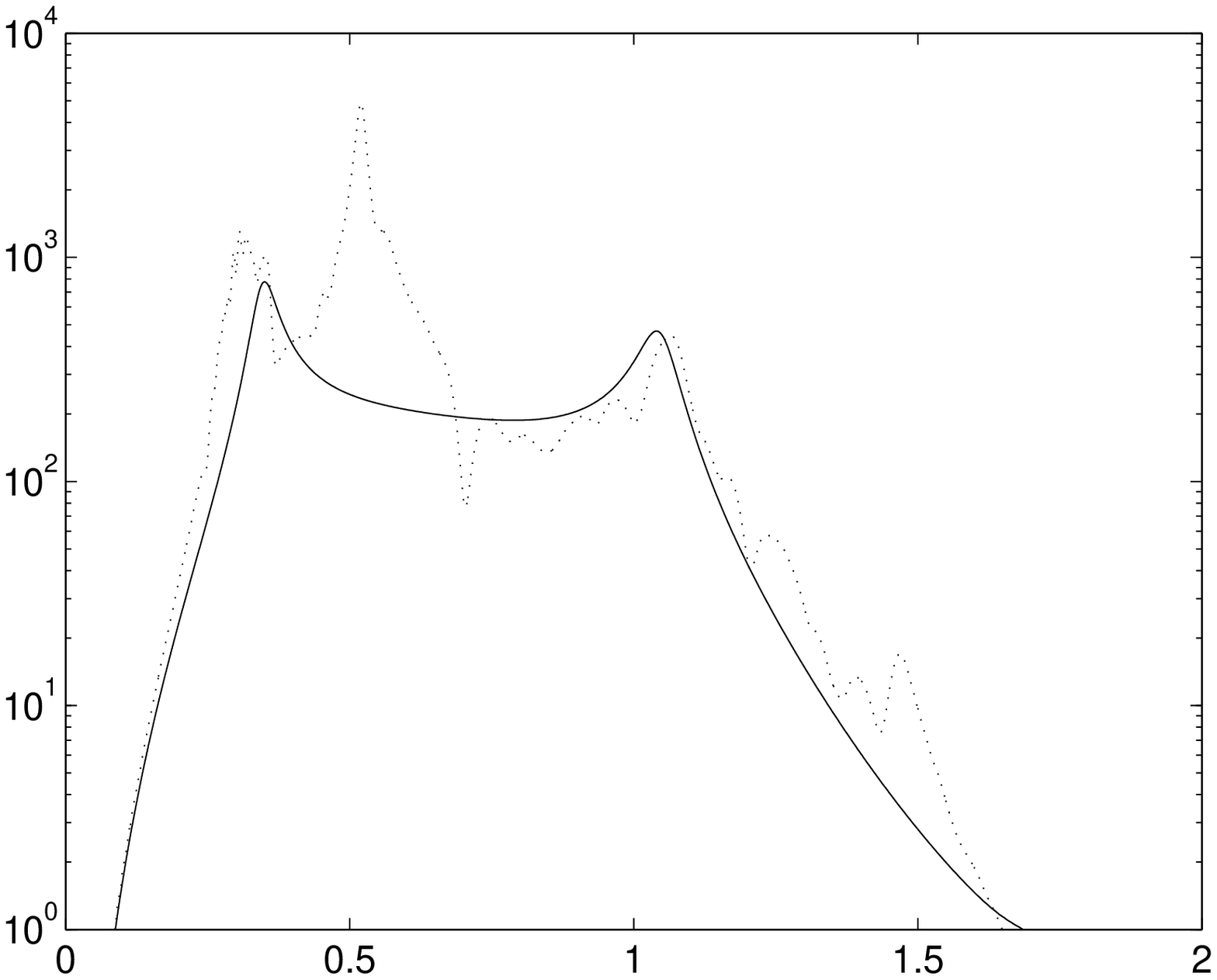}
\includegraphics[width=4.5cm]{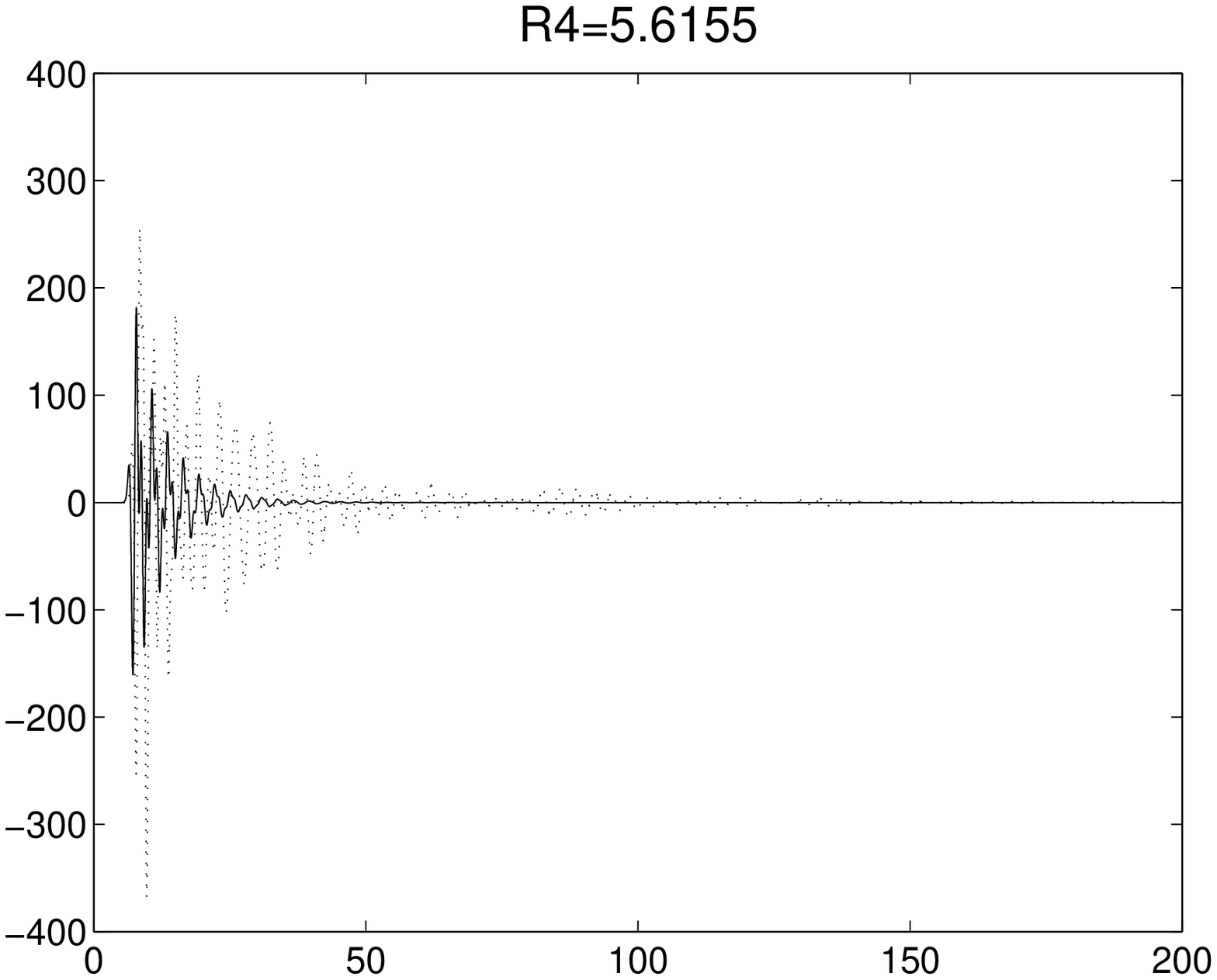}\hfill
\includegraphics[width=4.5cm]{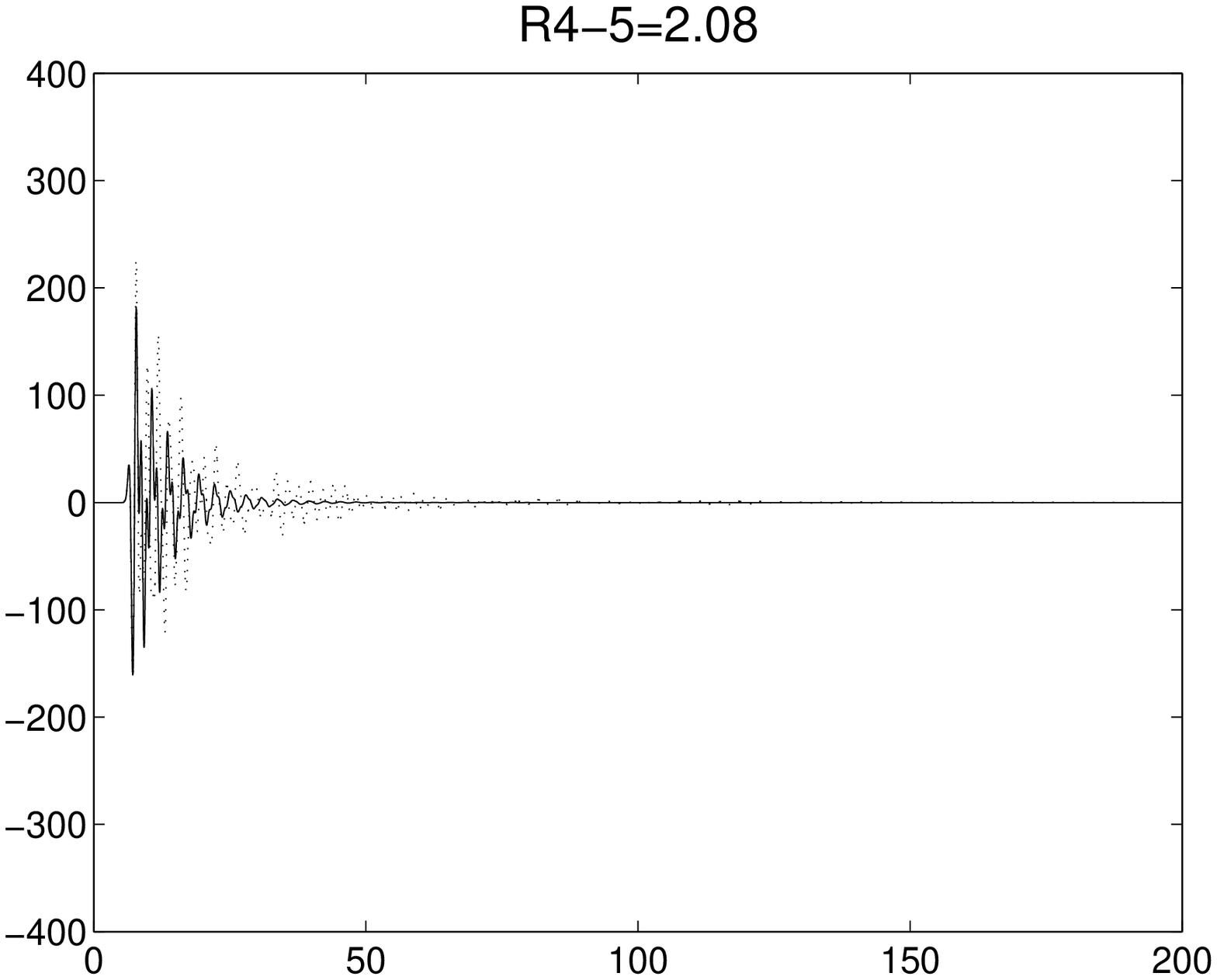}\hfill
\includegraphics[width=4.5cm]{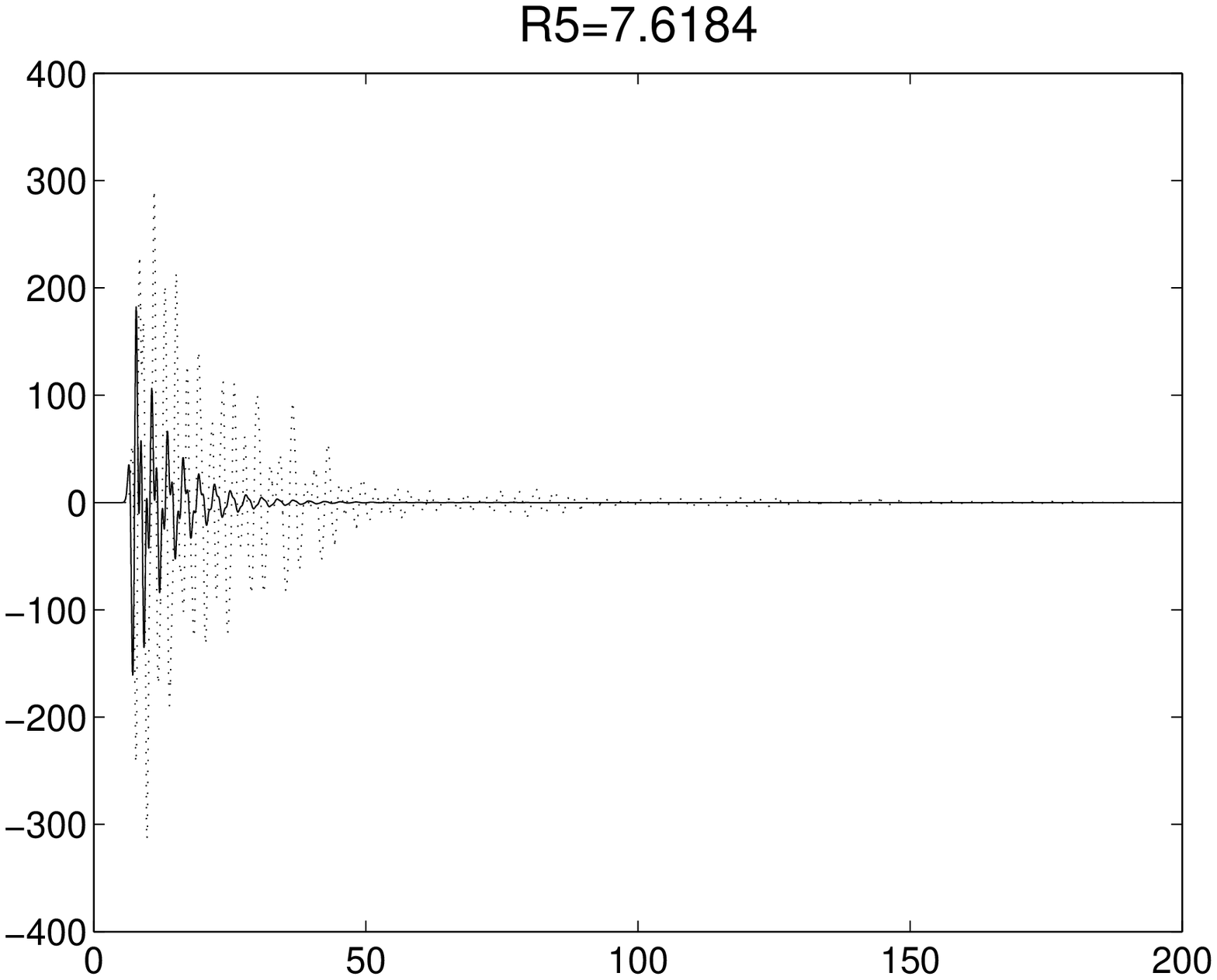}\hfill
\includegraphics[width=4.5cm]{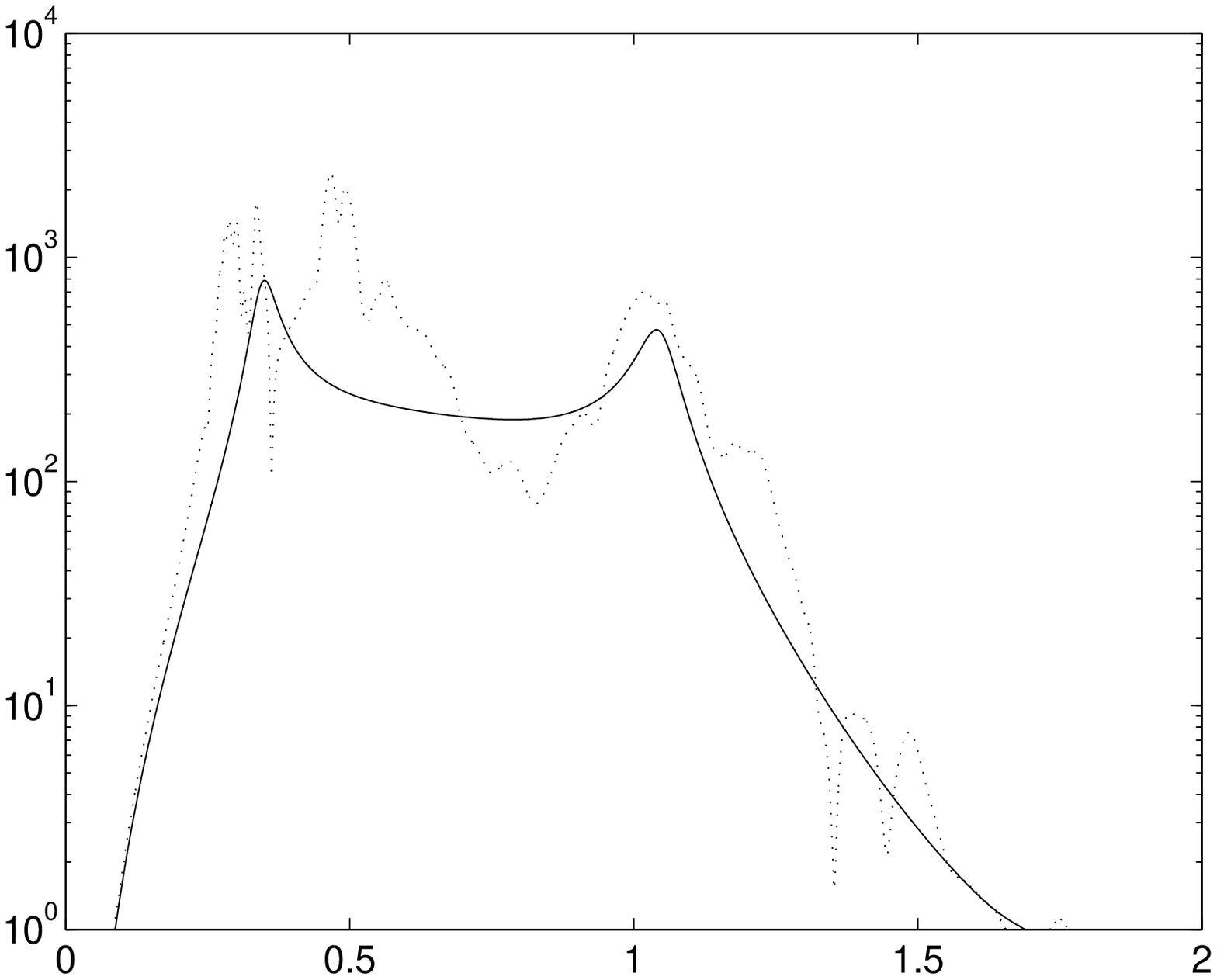}\hfill
\includegraphics[width=4.5cm]{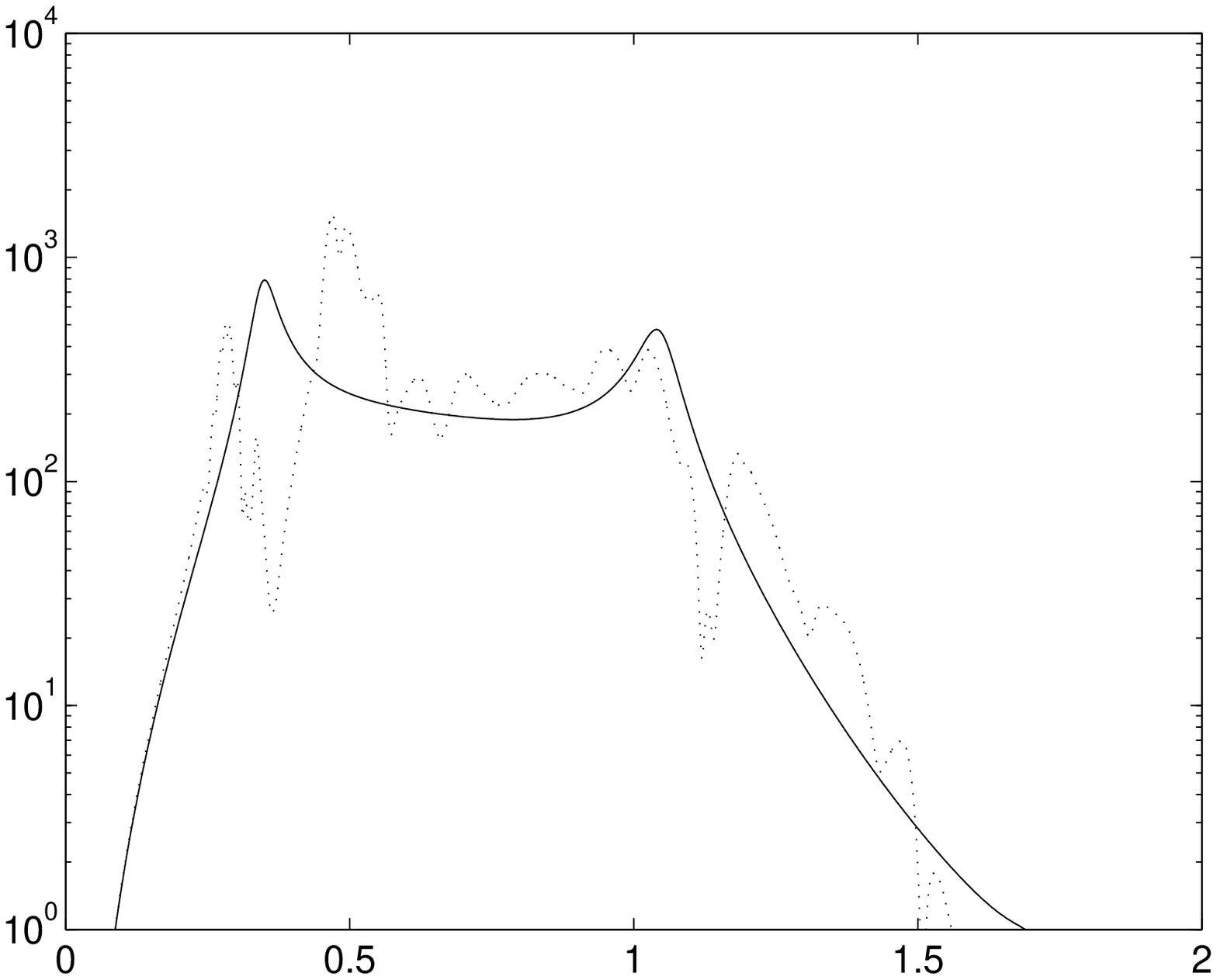}\hfill
\includegraphics[width=4.5cm]{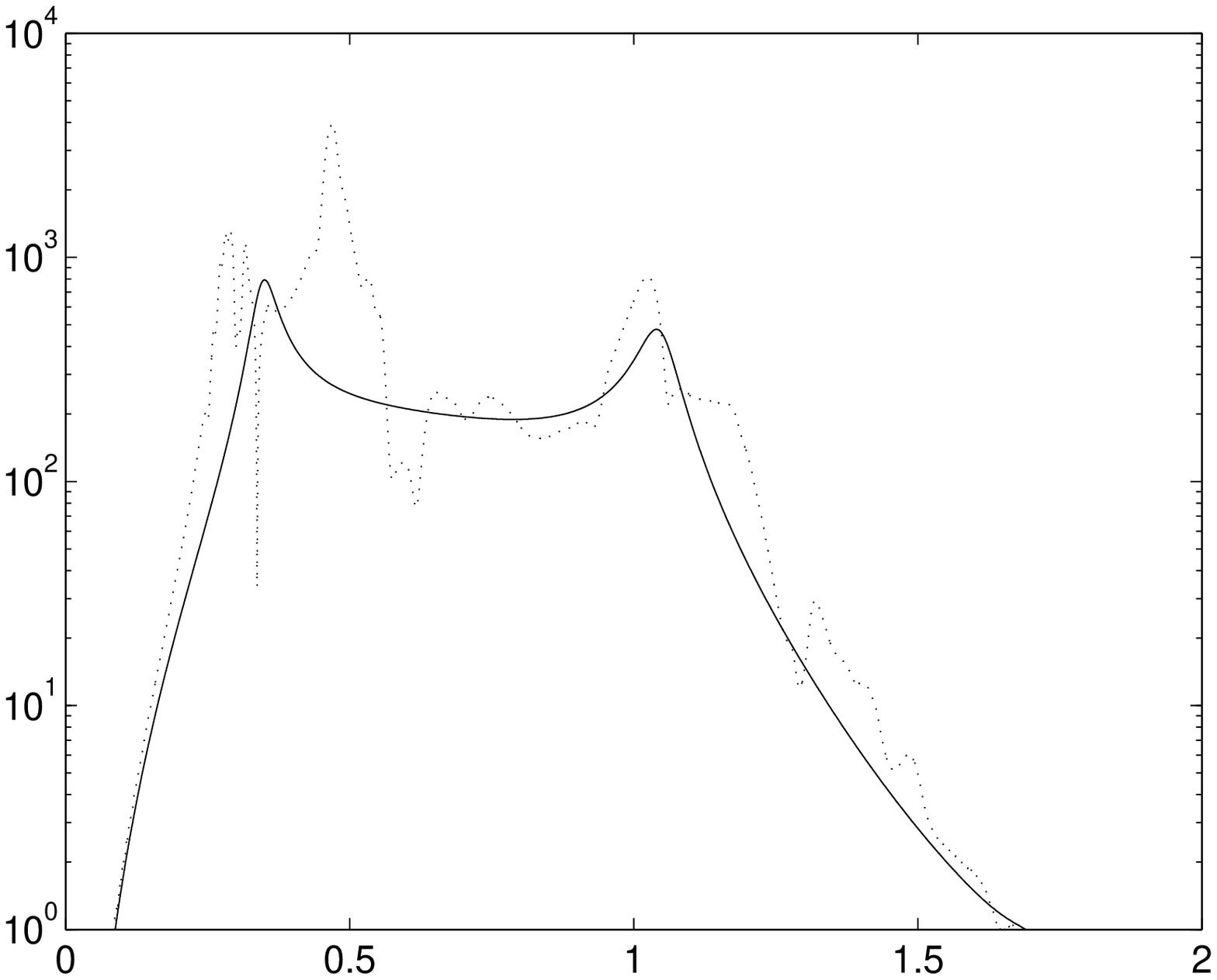}\hfill
\caption{Time records of total particle velocity for  `city' $C^{2}$ with ten blocks having different spacing $d_{j,j+1}^{1}$ ($j=1,2,...,9$). Each row of the figure depicts the particle velocity (in $s$): at the center of the top of the $j$-th block (left), the center of the ground segment between the $j$-th and the $j+1$-th block (middle) and at the top of the $j+1$-th block (right). The solid curves in all the subfigures represent the particle velocity at ground level in the absence of the blocks. The vulnerability indices $R_{j}$ at the top of the $j$-th block and $R_{j,j+1}$  on the ground between the $j$-th and the $j+1$-th block, are indicated at the top of each subfigure. The abscissas designate time, and range from $0$ to $200s$. Note that the scales of the ordinates do not vary from one subfigure to another. Below each time record appears the modulus of the velocity spectrum of the previous quantity. The abscissas designate frequency, and range from $0$ to $2Hz$.}
\label{TimeC2}
\end{figure}

\begin{figure}[H]
\includegraphics[width=4.5cm]{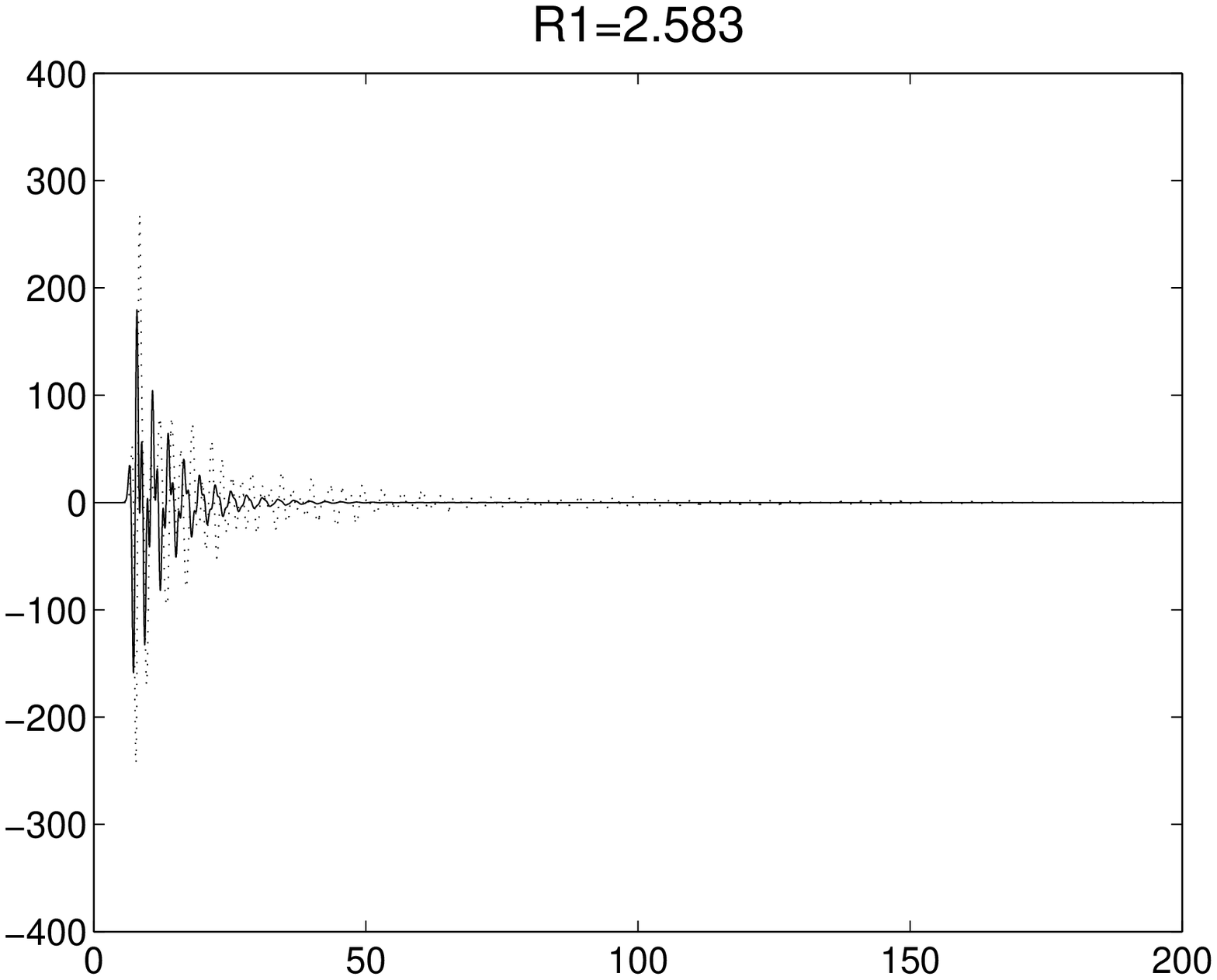}\hfill
\includegraphics[width=4.5cm]{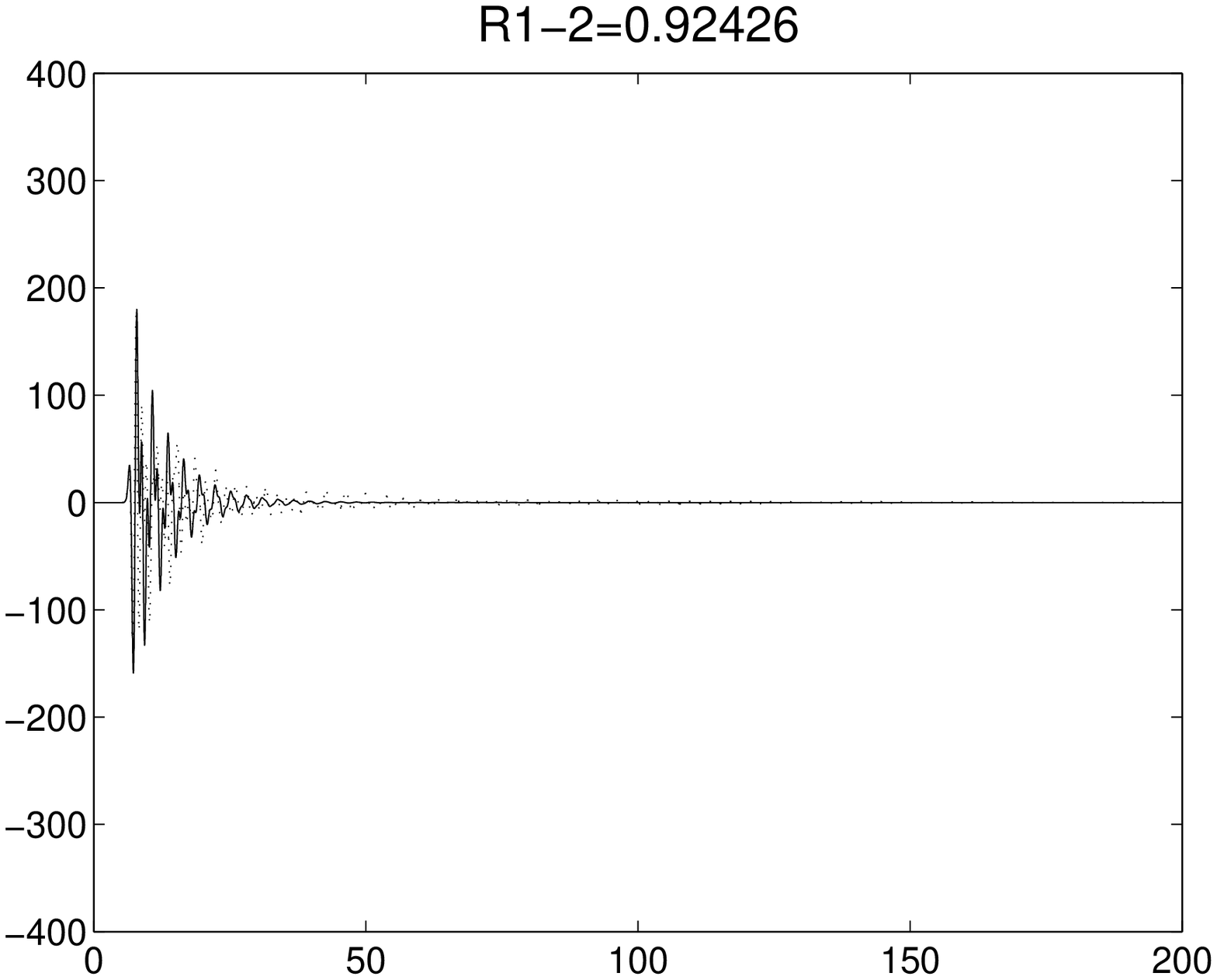}\hfill
\includegraphics[width=4.5cm]{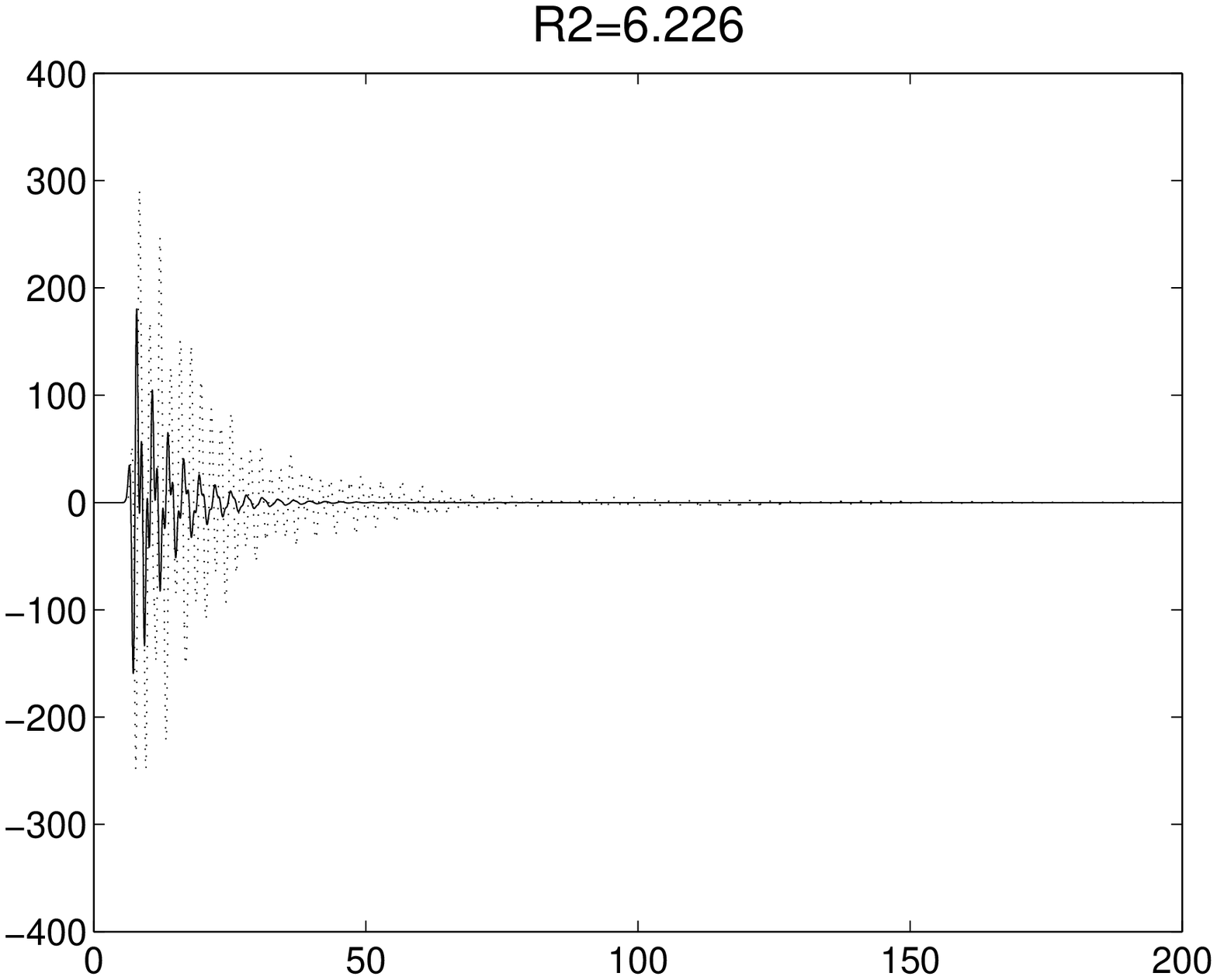}
\includegraphics[width=4.5cm]{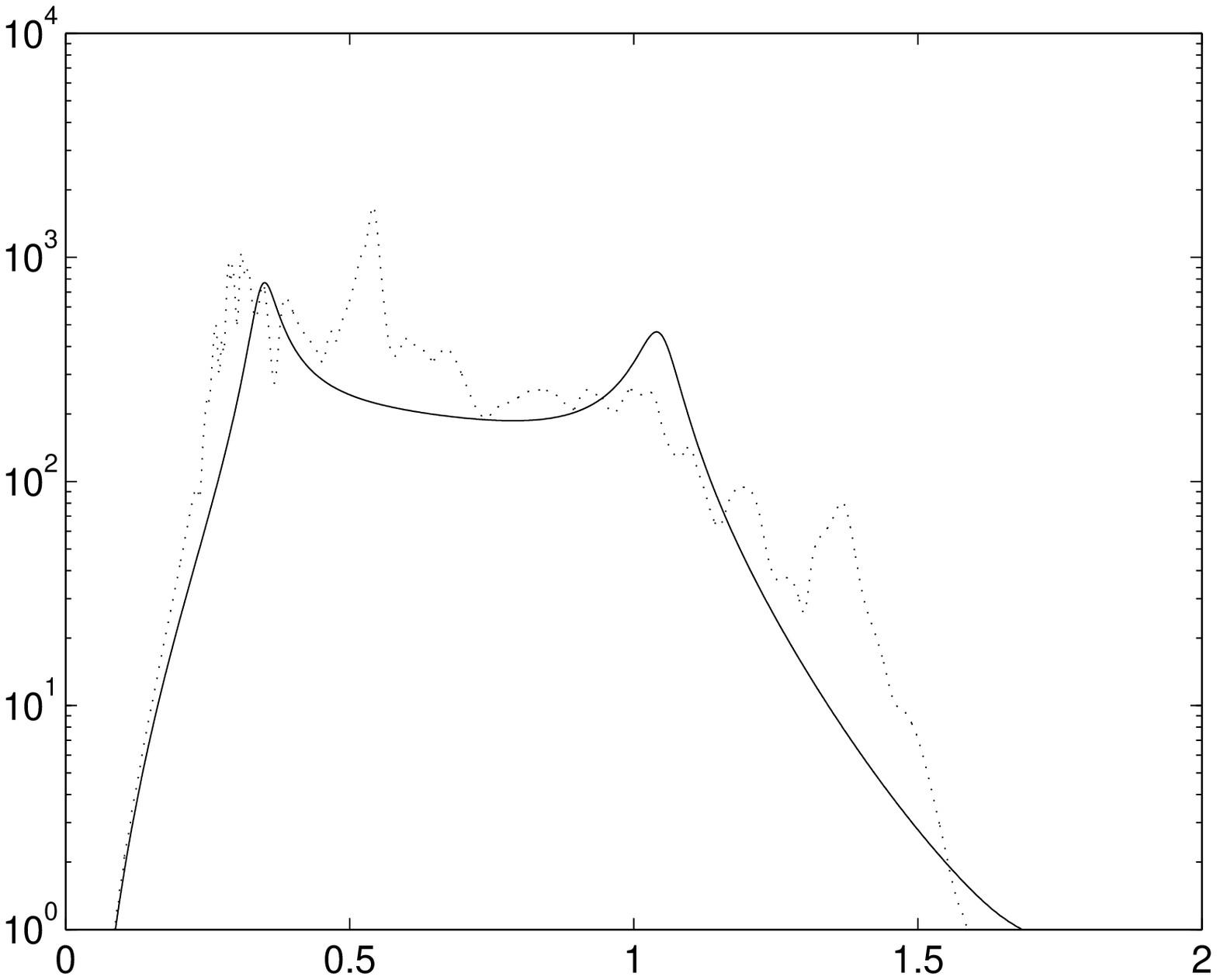}\hfill
\includegraphics[width=4.5cm]{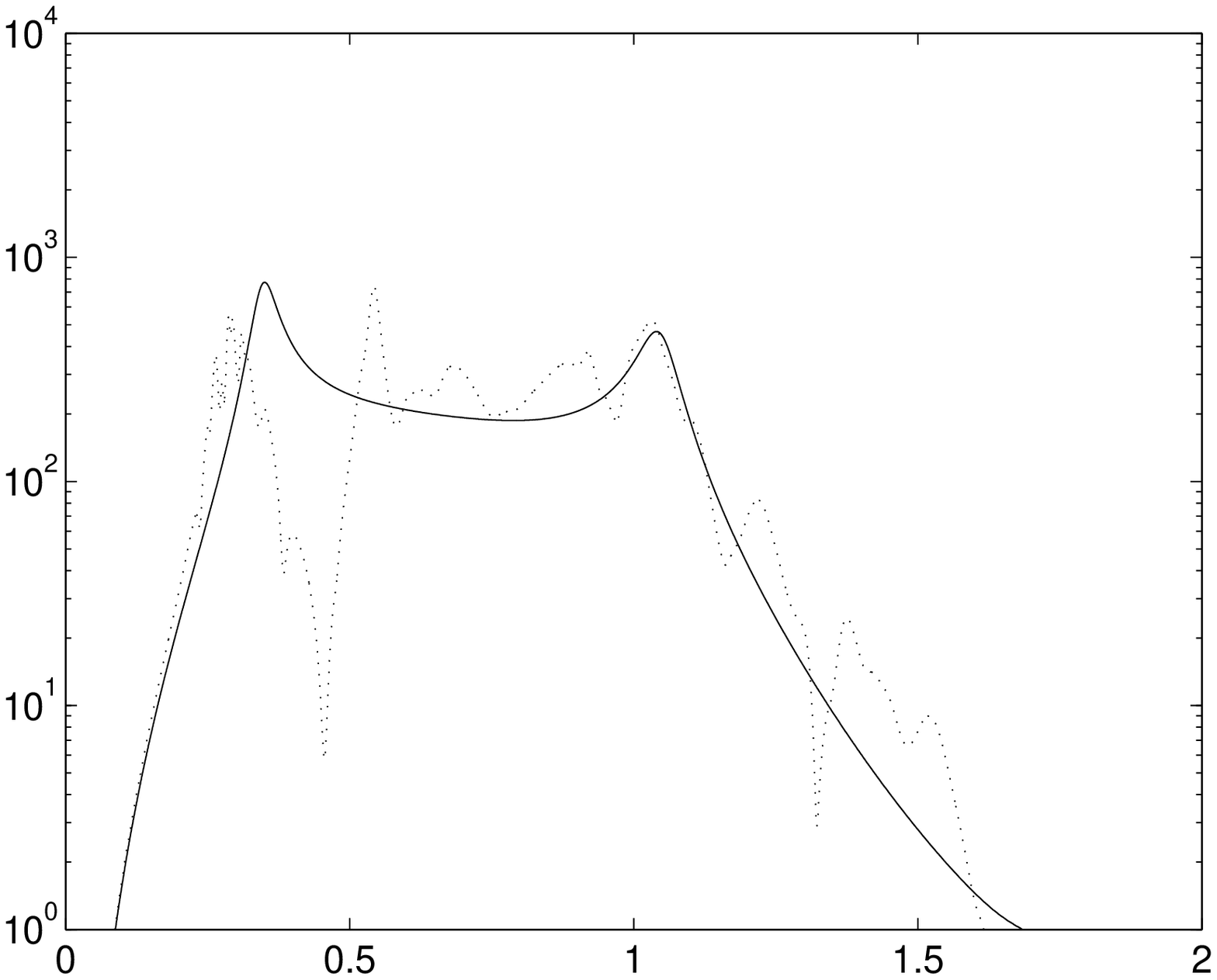}\hfill
\includegraphics[width=4.5cm]{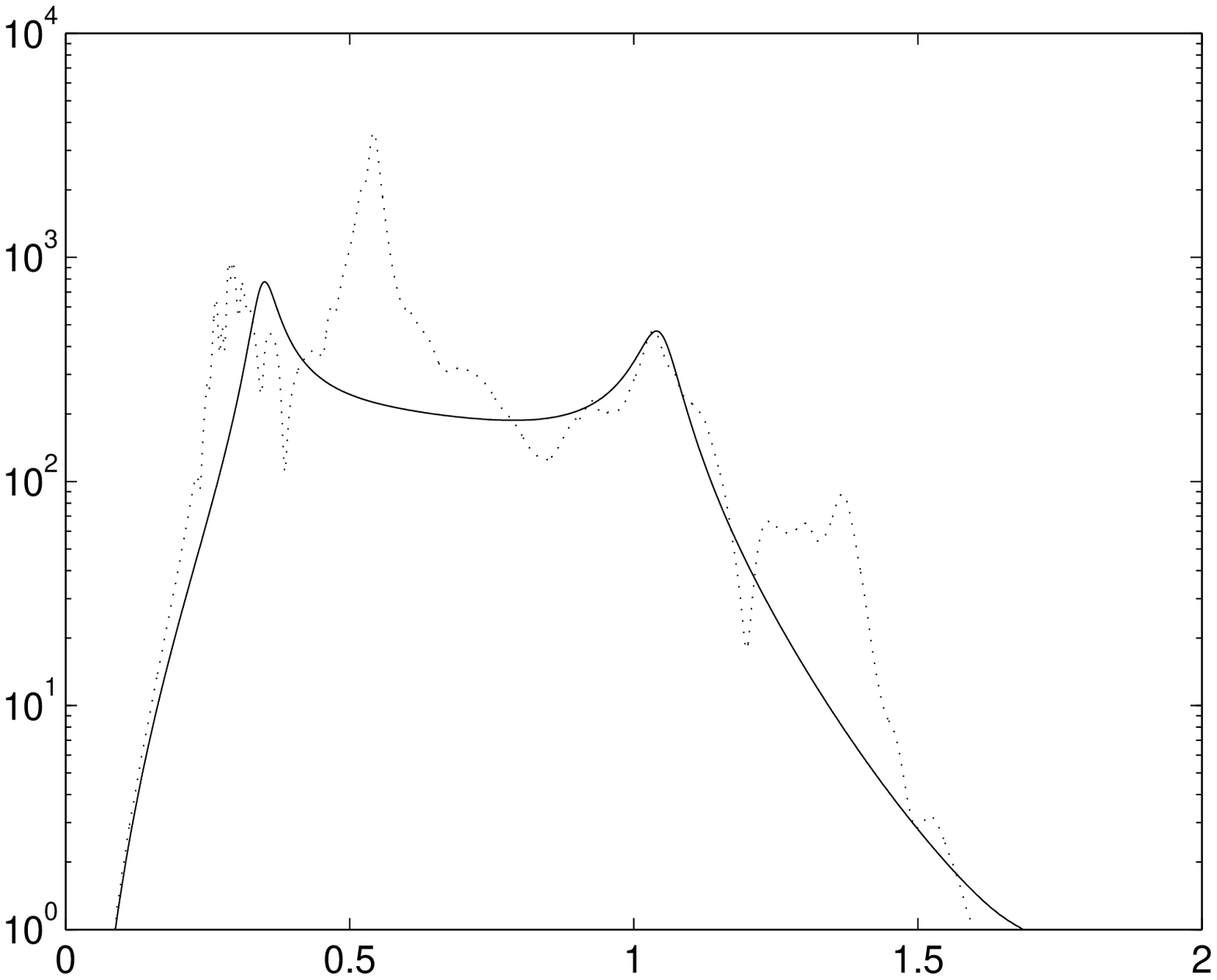}
\includegraphics[width=4.5cm]{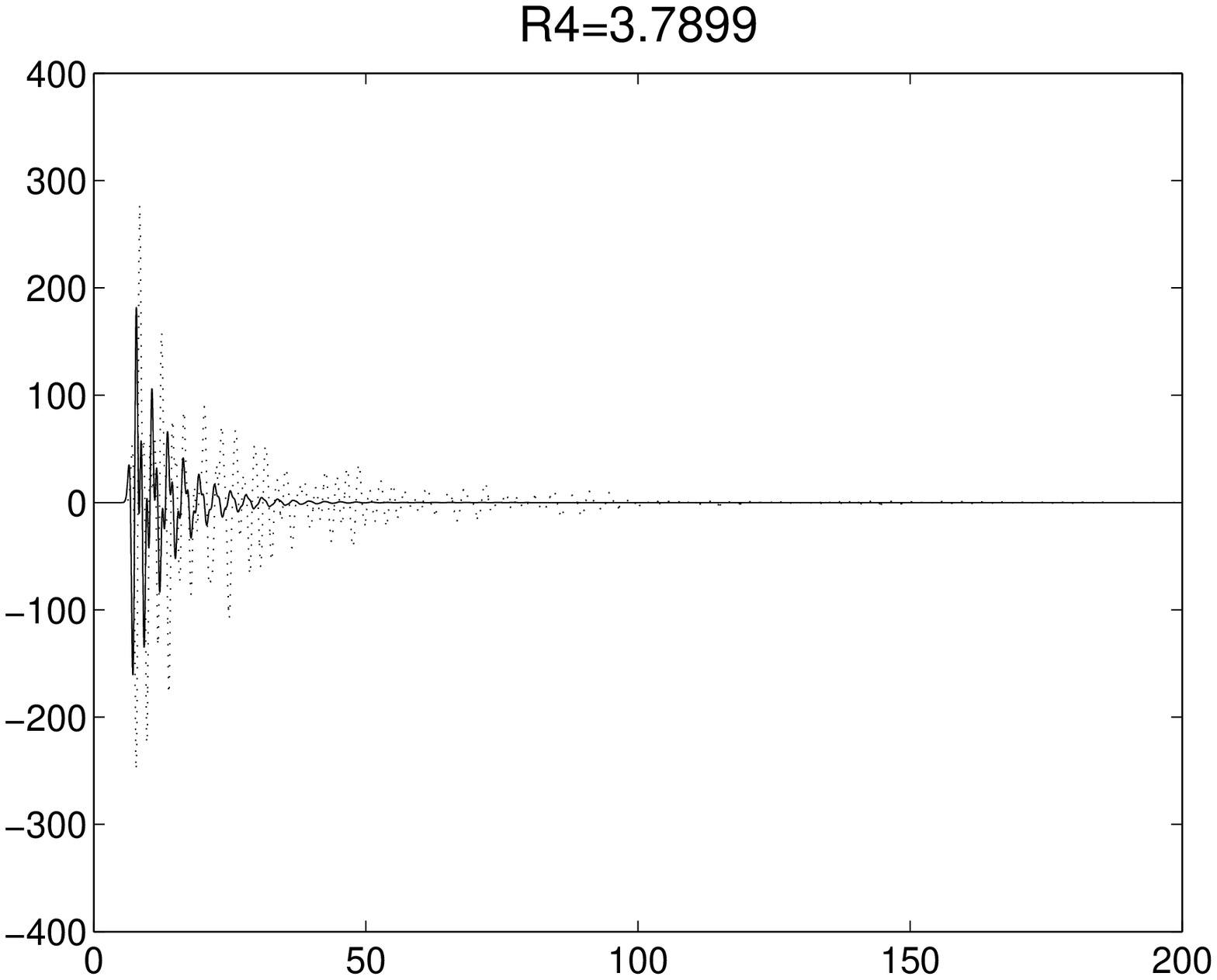}\hfill
\includegraphics[width=4.5cm]{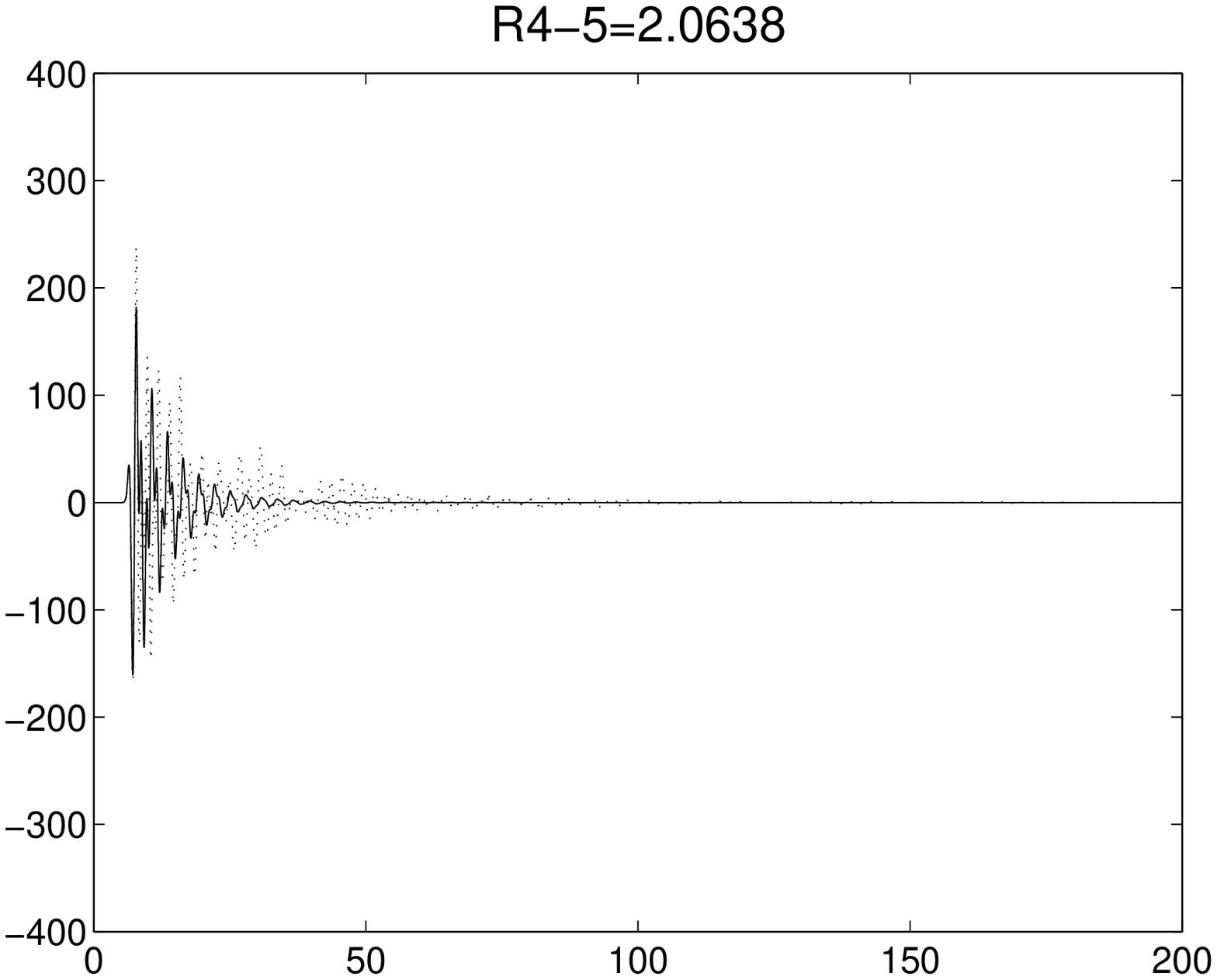}\hfill
\includegraphics[width=4.5cm]{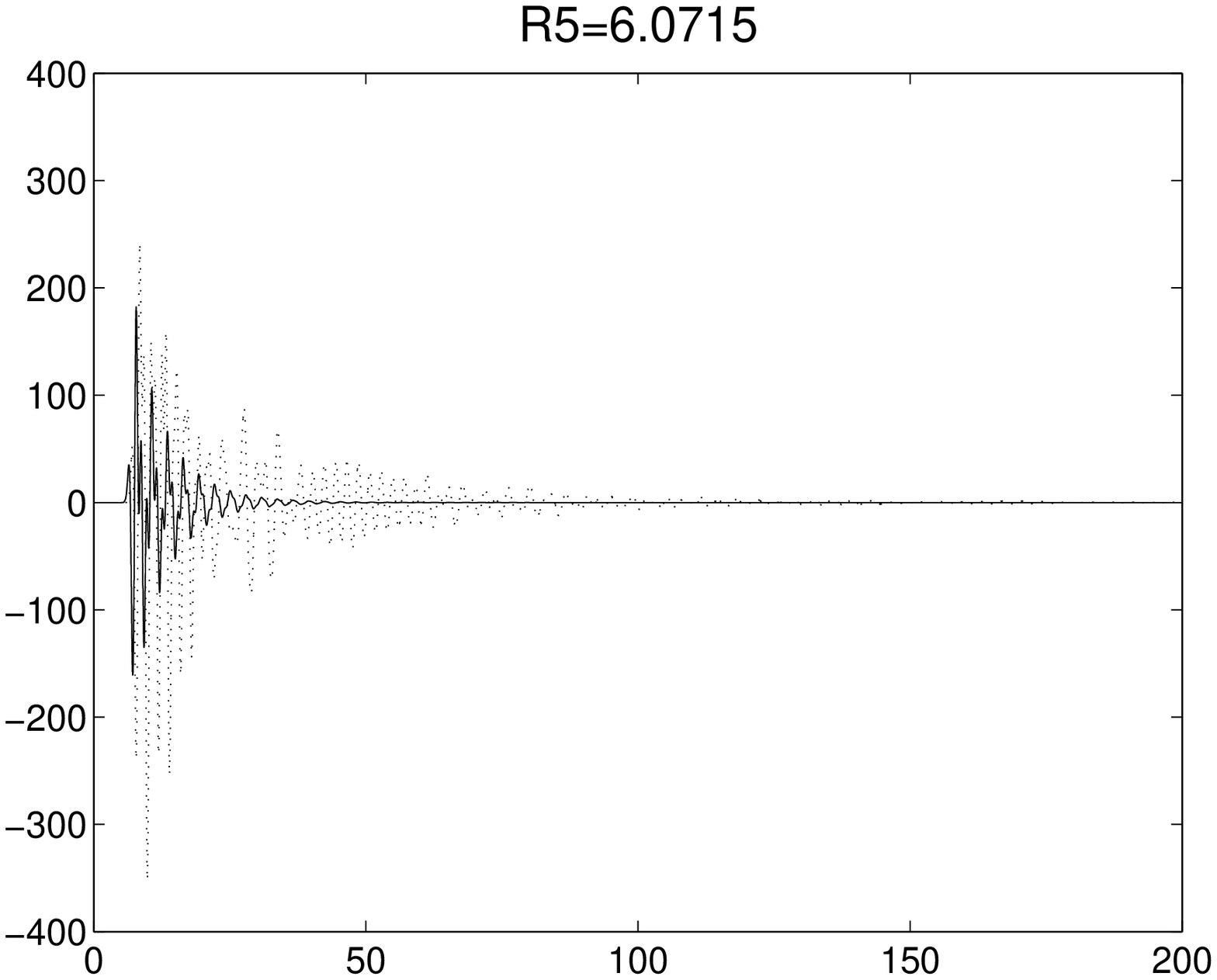}\hfill
\includegraphics[width=4.5cm]{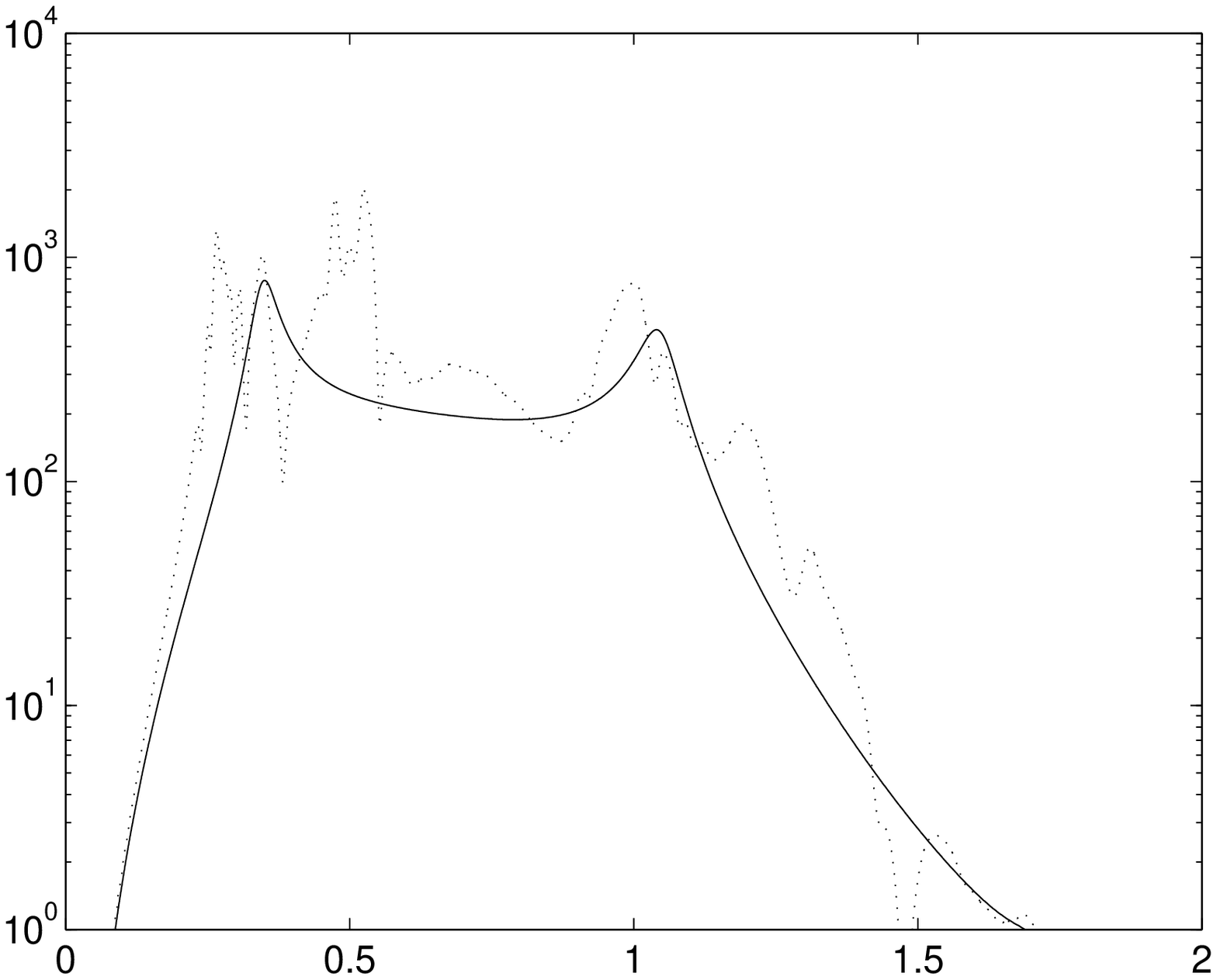}\hfill
\includegraphics[width=4.5cm]{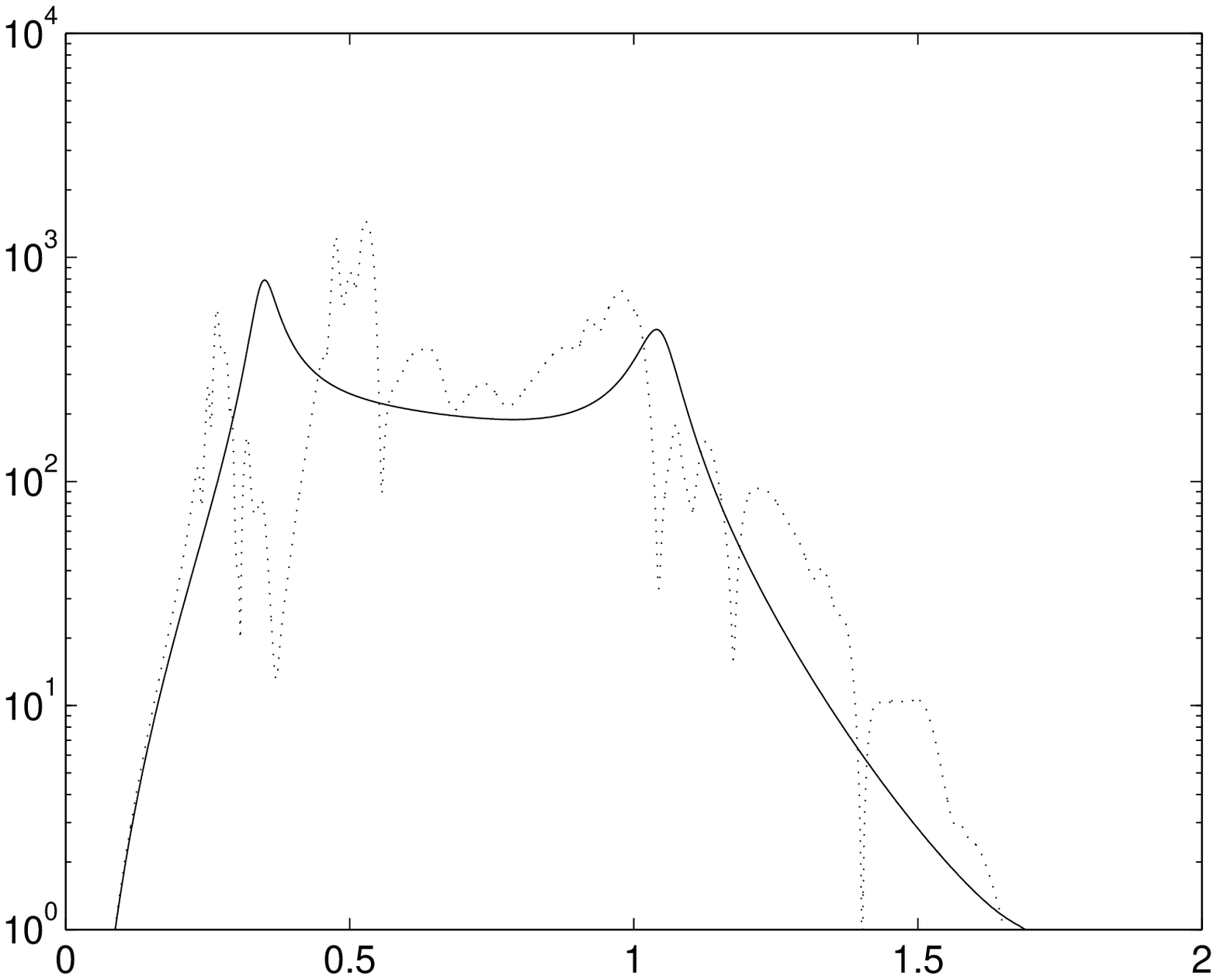}\hfill
\includegraphics[width=4.5cm]{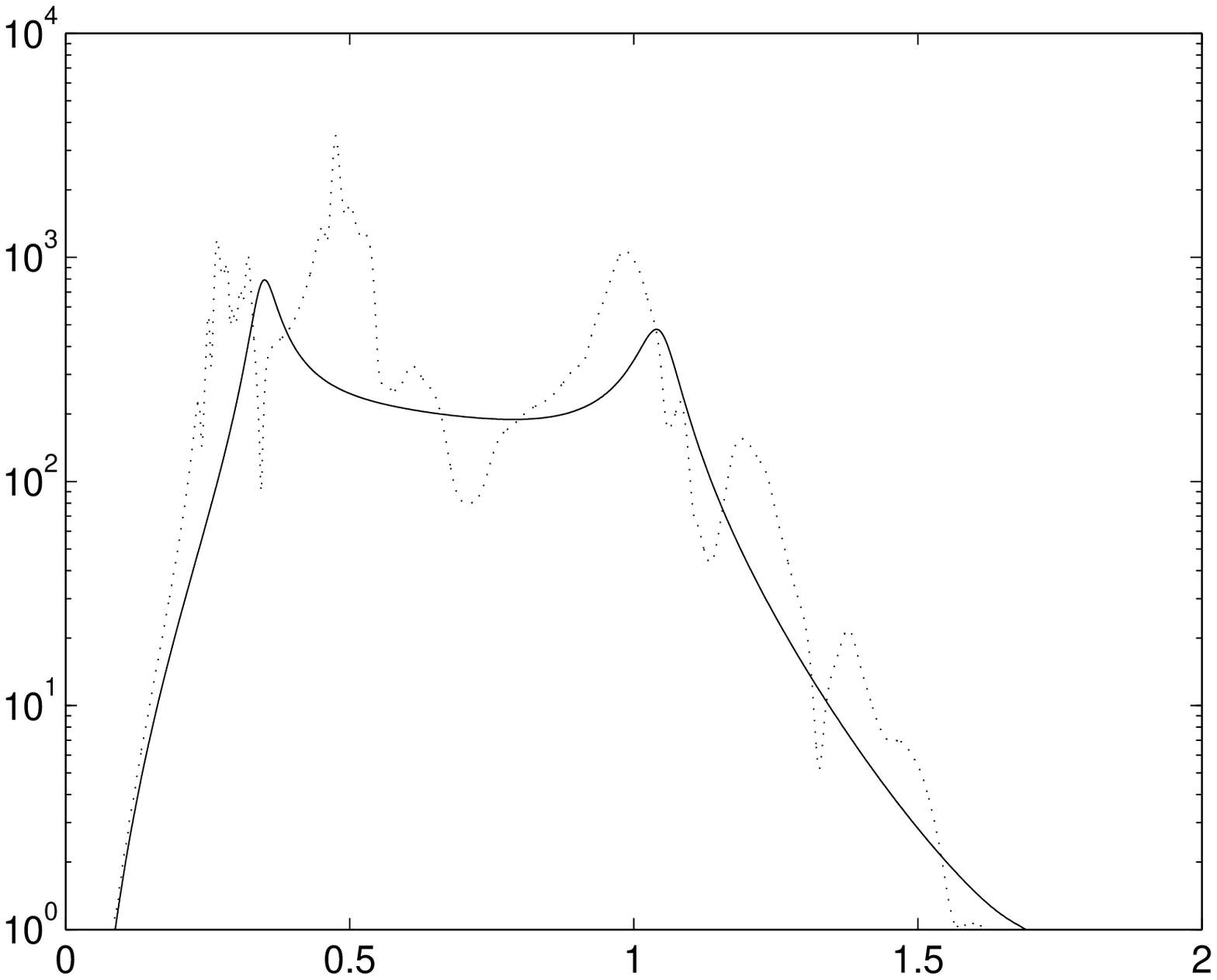}\hfill
\caption{Time records of total particle velocity for  `city' $C^{3}$ with ten blocks having different spacing $d_{j,j+1}^{2}$ ($j=1,2,...,9$). Each row of the figure depicts the particle velocity (in $s$): at the center of the top of the $j$-th block (left), the center of the ground segment between the $j$-th and the $j+1$-th block (middle) and the $j+1$-th block (right). The solid curves in all the subfigures represent the particle velocity at ground level in absence of blocks. The vulnerability indices $R_{j}$ at the top of the $j$-th block and $R_{j,j+1}$  on the ground between the $j$-th and the $j+1$-th block, are indicated at the top of each subfigure. The abscissas designate time, and range from $0$ to $200s$. Note that the scales of the ordinates do not vary from one subfigure to another. Below each time record appears the velocity spectrum of the previous quantity. The abscissas designate frequency, and range from $0$ to $2Hz$.}
\label{TimeC3}
\end{figure}

The time records and spectra, presented in Figs. \ref{CompV6}-\ref{TimeC3} call for the following comments. 
\begin{itemize}
\item Fig. \ref{CompV6}  exhibits the time history of the velocity at the summit of  building no. 6, for both the dissipative and non-dissipative cases. The durations are much longer when there is no dissipation. Nevertheless, in the dissipative case, beatings are observed and  the duration of motion is of the order of 2 min. 
\item Fig. \ref{TimeC1} contains the time histories of the velocity, and the velocity spectra at the summits and on the ground in between blocks for city $C^{1}$. 
\item Fig. \ref{TimeC2} contains the time histories of the velocity, and the velocity spectra at the summits and on the ground in between blocks for city $C^{2}$. 
\item Fig. \ref{TimeC3} contains the time histories of the velocity, and the velocity spectra at the summits and on the ground in between blocks for city $C^{3}$.  
\end{itemize}
The results in terms of the vulnerability indices are given in Table \ref{t2}.
\section{Discussion}
This work was initiated in \cite{Wir2003}, in which the authors considered the soft layer and blocks to be occupied by {\it non-dissipative media}.

The time histories in Figs. \ref{TimeC1}-\ref{TimeC3} herein, relative to {\it dissipative} blocks and a {\it dissipative} soft layer, again exhibit  amplification, beating phenomena, long codas and spatial variability of response in all three cities. The spectra exhibit so-called 'splittings' \cite{Wir1996}, which are responsible for beating phenomena and part of the duration. The durations (time during the which amplitude is more than $0.1$ of the maximal amplitude) are not notably different for the three cities (i.e., maximal duration of around $120s$) for the three configurations. The entries in Table \ref{t2} indicate that the vulnerability indices of the blocks are significantly-large even in the case of dissipative media. The spectra are more irregular (i.e., the number of peaks is larger) when the average separation of the blocks is smaller. This may constitute an indication of greater structure-soil-structure and/or greater block-to-block interaction responsible for more efficient excitation of quasi Love modes (these are neither the rigid base block modes, nor the Love modes in the absence of the buildings, but a combination of the two \cite{wiko93}).
\section{Conclusion}
The traditional \cite{baee88}, \cite{chba94} (and even quite recent \cite{fasu94}, \cite{ol00-7}, \cite{sedu00}) choice, concerning the analysis of earthquakes in cities built on soft soil as Nice and Mexico City, has been to introduce complexity into the substratum while leaving the free surface flat or with large-scale topographic features \cite{sedu00}. The buildings are not included in such  models and their response is treated separately, if at all, using the flat ground motion as the input. This has led to response predictions that are more in line with what has actually been observed during tremors in a variety of cities (Nice, Los Angeles, Mexico City, etc.). Nevertheless, one or several of the features, namely duration, peak velocities and spatial variability, of observed response, differ from those of the predictions. This is possibly due to the fact that the fine structure of the substratum is usually unknown. Another plausible hypothesis is that small-scale irregularities on the free surface and on interface between the foundations and the soft soil, introduced by the existence of buildings in a city, may contribute significantly to the overall motion of the site.

This hypothesis was shown to be tenable for: i) a non-dissipative basement  underlying a city of ten, non-equispaced, non-equally sized, homogenized, non-dissipative blocks in \cite{Wir2003}, and ii)  a dissipative basement underlying a city with an infinite number of equispaced, identical, dissipative blocks in \cite{Wir1996}. Herein, we again considered cities with ten non-equispaced non-equally sized, homogenized blocks, but the latter, as well as the underlying layer, were considered to be dissipative. 

We found that the introduction of dissipation does not eliminate the previously-found anomalous effects arising from the presence of the blocks, which take the form of peak ground and block motion amplifications, beatings, and long codas. As expected, the durations turned out to be smaller than in the non-dissipative case, but the vulnerability indices remained significantly large. When the spacing between blocks was reduced, we found some evidence of increased structure-soil-structure and block-to-block interactions, but, contrary to expectation, this did not have significant effects on either the duration of motion or on the peak amplification. 

These anomalous effects, as well as their spatial variability, are qualitatively the same as those observed in  Mexico City and underline the fact that the individual buildings or groups of buildings (blocks)  play a very active role in the overall motion of a city submitted to a seismic disturbance.

In the future, it will be necessary to examine to what extent the anomalous response is affected by the location and type of seismic source, as well as by the duration of the pulse radiated from this source (note that studies such as \cite{fasu94}, \cite{grwi04} take no account of the buildings of the city).  
\bibliography{biblio} 
\bibliographystyle{unsrt}
\newpage
\newpage
\newpage
\newpage
\newpage
\newpage
\newpage
\newpage
\begin{table}[htbp]
\begin{center}
\begin{tabular}{||c|c|c|c|c|c|c|c|c|c|c||}
\hline
$~$ & $B_{1}$ & $B_{2}$ & $B_{3}$ & $B_{4}$ & $B_{5}$ & $B_{6}$ & $B_{7}$ & $B_{8}$ & $B_{9}$ & $B_{10}$\\
\hline
\textbf{height }$h_{i}$ & 50 & 50 & 50 & 60 & 60 & 60 & 70 & 70 & 70 & 70\\
\hline
\textbf{width} $w_{i}$& 30 & 40 & 50 & 30 & 40 & 60 &30 &40 &50 &60 \\ 
\hline
\textbf{spacing} $d_{i,i+1}^{1}$& 70 & 90 & 60 & 80 & 100 & 60 &90 &80 &60 & \\
\hline
$d_{i,i+1}^{2}$& 30 & 40 & 30 & 40 & 50 & 30 &40 &40 &30 & \\
\hline
$d_{i,i+1}^{3}$& 10 & 20 & 10 & 20 & 30 & 10 &20 &20 &10 & \\
\hline
\end{tabular}
\caption{Geometrical parameters of the ten-block idealized cities. The units are meters}
\label{t1}
\end{center}
\end{table}
\begin{table}[htbp]
\begin{center}
\begin{tabular}{||c|c|c|c|c|c|c|c|c|c|c||}
\hline
 $~$ & $R_{1}$ & $R_{2}$ & $R_{3}$ & $R_{4}$ & $R_{5}$ & $R_{6}$ & $R_{7}$ & $R_{8}$ & $R_{9}$ & $R_{10}$\\
\hline
$\mathbf{C^{1}}$ & \bf{9.2} & \bf{10.2} & \bf{7.5} & \bf{12.6} & \bf{9.4} & \bf{6.1} & \bf{10.9} & \bf{7.9} & \bf{6.9} & \bf{6.5} \\ 
\hline
$C^{1}$ & 7.5 & 7.1 & 4.0 & 6.7 & 5.9 & 3.2 & 6.0 & 5.4 & 4.0 & 3.3 \\ 
\hline
$C^{2}$ & 4.2 & 9.5 & 5.9 & 5.6  & 7.6 & 4.0 & 5.0 & 6.6 & 4.0 & 2.7 \\ 
\hline
$C^{3}$ & 2.6 & 6.2 & 4.9 & 3.8 &  6.1 & 4.8 & 5.7 & 7.5 & 4.6 & 2.7 \\ 
\hline
\end{tabular}
\caption{Vulnerability indices. (bold for non-dissipative media)}
\label{t2}
\end{center}
\end{table}
\end{document}